\documentclass[a4paper,10pt]{article}
\newcommand{\head}[1]{\textnormal{\textbf{#1}}}
\usepackage[onehalfspacing]{setspace}
\usepackage{amsmath}
\usepackage{amssymb}
\usepackage{hyperref}
\usepackage{fontenc}
\usepackage{inputenc}
\usepackage{authblk}
\usepackage{color}
\usepackage{graphicx}
\usepackage{cite}
\usepackage{slashed}
\hypersetup{colorlinks=true,linkcolor=magenta,linktoc=page,citecolor=blue}
\usepackage{tikz}
\usepackage{multirow}
\usepackage{hhline}
\usepackage[top=1.5cm,bottom=2.5cm,right=2.54cm,left=2.54cm]{geometry}
\definecolor{c1}{rgb}{0.5,0,1}
\colorlet{aqua}{-red!75}
\numberwithin{equation}{section}
\begin{document}

	\begin{flushright}
		IPM/P-2016/006 \\
	\end{flushright}
	
	\vspace*{20mm}
	\begin{center}
		{\Large {\bf On the Monogamy of Holographic $n$-partite Information}\\}
		
		\vspace*{15mm} \vspace*{1mm} {S. Mirabi${}^{a}$,  M. Reza Tanhayi${}^{a,b}$ and R. Vazirian${}^a$}
		
		\vspace*{1cm}
		
		{${}^a$Department of Physics, Faculty of Basic Science, Islamic Azad
			University Central Tehran Branch (IAUCTB), P.O. Box 14676-86831,
			Tehran, Iran\\
			${}^b$School of Physics, Institute for Research in Fundamental Sciences
			(IPM) P.O. Box 19395-5531, Tehran, Iran}
		
		\vspace*{0.5cm}
		{E-mail: {\tt mtanhayi@ipm.ir,\,\,\,r.vazirian@srbiau.ac.ir}}
		
		\vspace*{1cm}
	\end{center}
	
	\begin{abstract}
		We investigate the monogamy of holographic $n$-partite information for a system consisting of $n$ disjoint parallel strips with the same width and separation in AdS and AdS black brane geometries. More precisely, we study the sign of this quantity, \emph{e.g.} for $n=4, 5$, in various dimensions and for different parameters. Our results show that for quantum field theories with holographic duals, the holographic 4-partite information is always positive and the sign of holographic 5-partite information is found to be negative in the dual strongly coupled $1+1$ dimensional CFT. This latter result indicates that the holographic 4-partite information is monogamous. We also find the critical points corresponding to the possible phase transitions of these quantities.
	\end{abstract}
	\newpage
	
	
	\section{Introduction}
	
	Entanglement entropy is one of the important non-local quantities
	which measures the quantum entanglement between different degrees
	of freedom of a quantum mechanical
	system\cite{Horodecki:2009zz,Casini:2009sr}. Similar to other
	non-local quantities, \emph{e.g.} Wilson loop and correlation
	functions, entanglement entropy can be used to classify the
	various quantum phase transitions and critical
	points\cite{Vidal:2002rm}.  In the context of quantum field theory
	there is a straightforward instruction to define this quantity
	(for example, see \cite{Callan:1994py,Calabrese:2009qy}). Consider
	a constant time slice in a $d$ dimensional quantum field theory
	and divide it into two spatial regions $A$ and $\bar{A}$ where
	they are complement to each other. For a local quantum field
	theory this geometrical division leads to a specific partitioning
	of the corresponding total Hilbert space as
	$\mathcal{H}=\mathcal{H}_A\otimes\mathcal{H}_{\bar{A}}$. Now the
	reduced density matrix for region $A$ can be computed by
	integrating out the degrees of freedom that live in ${\bar{A}}$,
	\emph{i.e.} $\rho_A={\rm Tr}_{\bar{A}}\;\rho$ where $\rho$ is the
	total density matrix. The entanglement entropy is given by the Von
	Neumann entropy for this reduced density matrix, \emph{i.e.}
	$S_A=-{\rm Tr}_{A}\;\rho_A \log \rho_A$.
	
	Entanglement entropy for local quantum field theories is infinite
	such that for $d>2$ the coefficient of the divergent term is
	proportional to the area of the entangling surface
	\cite{Bombelli:1986rw, Srednicki:1993im}
	\begin{align}
		S_{\rm{EE}}\propto \frac{\mathcal{A}_A}{\epsilon^{d-2}}+\cdots,
	\end{align}
	where $\epsilon$ is the UV cut-off. This behavior is the
	celebrated area law which is due to the infinite correlations
	between degrees of freedom near the boundary of entangling
	surface. Beside the elegant role of entanglement entropy in
	various physical contexts, \emph{e.g.} quantum information theory
	and black hole physics, it has some features which are less
	pleasant. Appearance of the UV cut-off in the expression of
	entanglement entropy makes it a non-universal quantity. Also, this
	quantity for a single entangling region cannot demonstrate all the physical content of a field
	theory, for example in a two dimensional conformal field theory
	(CFT) entanglement entropy for a single interval only depends on the central charge of
	the theory and other aspects of CFT are
	absent\cite{Calabrese:2004eu}. According to these features, it
	seems natural to search for other useful quantities to improve our
	knowledge of the Hilbert space of a quantum system.
	
	In the context of quantum information theory many quantities were
	defined to overcome the shortcomings that we encountered using
	entanglement entropy, \emph{e.g.} mutual and tripartite
	information. Mutual information for a system which has two
	disjoint parts is given by\cite{Casini:2008wt}
	\begin{align}\label{HMI}
		I^{[2]}(\equiv I)(A:B)=S(A)+S(B)-S(A\cup B),
	\end{align}
	where $S(A\cup B)$ is the entanglement entropy for the union
	of the two entangling regions. This is a finite quantity and
	quantifies the amount of entanglement (information) which is
	shared between two subsystems. In
	\cite{Calabrese:2009ez,Calabrese:2010he} it was shown that mutual
	information in a two dimensional CFT depends explicitly on the
	full operator content of the theory and not only the central
	charge. Another interesting feature of the mutual information is
	its positivity which is due to the strong subadditivity property
	for entanglement entropy.
	
	In addition to the entanglement entropy and mutual information, another useful quantity which is defined for a
	system consisting of three disjoint parts is tripartite
	information. This quantity is defined as
	follows\cite{Casini:2008wt}
	\begin{align}\label{3par}
		{I^{[3]}}({A}:{B}:{C}) &= S({A}) + S({B}) + S({C}) - S({A} \cup {B}) - S({A} \cup {C}) \notag\\
		&- S({B} \cup {C}) + S({A} \cup {B} \cup {C}),
	\end{align}
	where $S(A\cup B\cup C)$ is the entanglement entropy for the
	union of three subsystems. Tripartite information is free of
	divergences and can be positive, negative or zero depending on the
	underlying field theory. It is important to mention that in spite
	of mutual information, tripartite information is finite even when
	the regions share boundaries. Also, in \cite{Kitaev:2005dm} it was
	shown that it is related to the topological entanglement entropy. According to the definition given by \eqref{3par}, the tripartite information can be written in terms of mutual information as follows
	\begin{align}\label{3par1}
		{I^{[3]}}({A}:{B}:{C}) &= I({A} : {B}) + I({A}: {C}) - I({A} : {B} \cup {C}).
	\end{align}
	Indeed as we explain below in the context of field theories with gravitational dual it was shown that the holographic tripartite information is always negative, \emph{i.e.} $ {I^{[3]}}({A}:{B}:{C}) <0$, which implies that the holographic mutual information is monogamous.\footnote{Any inequalities in the form of $F({A}:{B})+F({A}:{C})\leq F({A}:{B}U{C})$, are known as monogamy relations which are characteristic of measures of quantum entanglement. In the context of quantum information theory, the monogamy property is related to the security of quantum cryptography, since, unlike classical
		correlation, quantum entanglement is not a shareable resource. In other words, entangled correlations between
		$A$ and $B$ cannot be shared with a third system $C$ without spoiling the original entanglement \cite{Hayden:2011ag}.}

	Intuitively, by generalizing the previous instruction to a system consisting of $n$
	partitions, one can define a new quantity which is called
	$n$-partite information. Similar to mutual and tripartite
	information, the $n$-partite information is given by
	\cite{Hayden:2011ag}
	\begin{align}\label{npar}
		I^{[n]}(A_1:\, \cdots \, :A_n)=\sum_{i=1}^nS(A_i)-\sum_{i<j}^nS(A_i\cup A_j)+\cdots-(-1)^{n}S(A_1\cup\cdots\cup A_n),
	\end{align}
 where in terms of the mutual information, the $n$-partite information may be written as follows
 \begin{eqnarray}
 I^{[n]}(A_1:\, \cdots\, :A_n)&=&\sum_{i=2}^nI(A_1:A_i)-\sum_{i=2<j}^n
 I(A_1:\,A_i\cup A_j)+\sum_{i=2<j<k}^nI(A_1:\,A_i\cup A_j\cup A_k)-\cdots\cr &&\cr
 &+&(-1)^nI(A_1:\,A_2\cup A_2\cdots\cup A_n).
 \end{eqnarray}
	This specific combination of the entanglement entropies between
	different subsystems leads to a finite quantity. Moreover, this
	definition is such that for $n=1$, $n=2$ and $n=3$ reduces to
	entanglement entropy, mutual information and tripartite
	information respectively. Other quantities can be extracted for
	higher values of $n$, \emph{e.g.} 4-partite and 5-partite
	information for $n=4$ and $n=5$. Similar to mutual and tripartite
	information, these quantities may help us to investigate the
	operator content of the theory. According to equation \eqref{npar}, one can find a simple relation between
	the $n$-partite and $(n-1)$-partite information as follows
	\begin{align}\label{npar2}
		I^{[n]}(A_{\{i\}})=I^{[n-1]}(A_{\{1,\cdots,n-2\}}:\,A_{n-1})+
		I^{[n-1]}(A_{\{1,\cdots,n-2\}}:\,A_n)-I^{[n-1]}(A_{\{1,\cdots,n-2\}}:\,A_{n-1}\cup A_n).
	\end{align}
	The above relation shows that the sign of $n$-partite information constraints the monogamy of $(n-1)$-partite information such that for any $n$ when $I^{[n]}<0$ the $I^{[n-1]}$ becomes monogamous.

	Although the previously mentioned quantities have interesting
	features which are useful to explore the Hilbert space of a
	quantum field theory, computing these quantities is not an easy
	task. Specifically, analytic computation of entanglement entropy
	is only done in few cases, \emph{e.g.} a two dimensional conformal
	field theory. In \cite{Ryu:2006bv} the authors proposed that by
	virtue of the AdS/CFT correspondence
	\cite{Maldacena:1997re,GKP:1998,W:1998} one can address this
	difficulty. According to this correspondence, a strongly coupled
	CFT with large number of degrees of freedom is dual to a classical
	gravity on an asymptotically AdS background. Using the dual
	gravity description, the prescription for finding the holographic
	entanglement entropy (HEE) associated to a spatial region has a
	simple geometric meaning. Consider a $d$ dimensional CFT which
	lives on the boundary of an AdS$_{d+1}$ space-time. Then the
	entanglement entropy for a spatial region $A$ is given by the
	following expression
	\begin{align}\label{ee}
		S_{\rm{EE}}=\frac{\mathcal{A}_{\rm{min}}(\gamma_A)}{4G_N},
	\end{align}
	where $G_N$  is the Newton's constant in $(d+1)$ bulk dimensions and $\mathcal{A}_{\rm{min}}(\gamma_A)$ is the minimal area of a
	co-dimension two hypersurface $\gamma_A$ in the bulk such that on
	the boundary of space-time $\partial \gamma_A=\partial A$. This
	prescription has passed several basic checks (see
	\cite{Nishioka:2009un} for a review). Generalization of this
	proposal to time dependent geometries and studying the
	thermalization process have also been done in
	\cite{Hubeny:2007xt,AbajoArrastia:2010yt,Albash:2010mv,Baron:2012fv,Liu:2013iza,Liu:2013qca,
		Alishahiha:2014cwa,Fonda:2014ula,Kundu:2016cgh}.
	
	Using Ryu and Takayanagi (RT) proposal one can also find the HEE
	for the union of disjoint entangling
	regions\cite{Hubeny:2007re,Headrick:2010zt,Ben-Ami:2014gsa,Tonni:2010pv,Faulkner:2013yia}.
	Particularly, in \cite{Headrick:2010zt} it was shown that the
	holographic mutual information (HMI) undergoes a first order phase
	transition by changing the distance between two regions. This
	phase transition is due to the competition between two different
	RT surfaces for computing $S(A\cup B)$.
	It is important to mention that the resultant HMI is a positive
	quantity as expected. Also, generalizing this procedure to systems
	consisting of more subsystems is straightforward. In this context,
	holographic tripartite and $n$-partite information in static and
	time dependent geometries have been studied in
	\cite{Hayden:2011ag,Balasubramanian:2011at,Allais:2011ys,Alishahiha:2014jxa,Mozaffar:2015xue,Tanhayi:2015cax}
	which show some interesting features. Actually, holographic
	tripartite information is always non-positive which leads to the
	monogamy of HMI\cite{Hayden:2011ag}. Also, in
	\cite{Alishahiha:2014jxa} it was shown that in a specific limit,
	the holographic $n$-partite information has a definite sign,
	\emph{i.e.} it is positive (negative) for even (odd) $n$. This is
	in contrast to the field theory results where the sign of these
	quantities is not fixed and depends on the underlying QFT. It is
	believed that these behaviors are a reminiscent of field theories
	which have gravitational dual and also may help us to more
	investigate the gravity side. Indeed, there is a strong connection
	between the energy constraints in the bulk and the sign of these
	quantities in the dual CFT. As an explicit example, in
	\cite{Allais:2011ys} it was shown that relaxing the null energy
	condition on the gravity side makes the holographic
	tripartite information positive and hence the monogamy of HMI
	breaks down.
	
	The main aim of this paper is to investigate the behavior of
	holographic $n$-partite information, \emph{e.g.} $n=4$ and $n=5$
	in different gravity set-ups. This study helps us to further investigate the monogamy of tripartite and 4-partite information. As we mentioned, the analysis of
	\cite{Alishahiha:2014jxa} is only valid for a specific limit
	where all of the entangling regions length $\ell$ and the separation
	between them $h$ are equal and also $h \ll \ell$. The authors have
	shown that assuming these conditions simplifies the expression for
	holographic $n$-partite information significantly and its sign
	becomes fixed. In the following sections, we will relax the $h \ll
	\ell$ condition and study the behavior of holographic $n$-partite
	information.
	
	The paper is organized as follows. In the next section, we will investigate the holographic $n$-partite information, specifically four and five partite information will be considered. In section three, we will study those quantities in AdS black brane (AdS-BH) background. Section four is devoted to conclusions and discussions. In the appendix, within a brief review, we will recall holographic entanglement entropy for a strip.
	
	\section{Holographic $n$-partite information in AdS geometry}
	
	For an entangling region, the entanglement entropy contains a short-distance divergence which is proportional to its area in the boundary. This makes the entanglement entropy to be scheme-dependent. However, for two disjoint systems one can introduce the mutual information as (\ref{HMI}) which is a finite and scheme-independent quantity. From the subadditivity property of entanglement entropy, it is proved that the mutual information is always non-negative quantity, $I({A}:{B})\geq 0$. On the other hand, for three disjoint regions say as ${A}$, ${B}$ and ${C}$, one can deal with tripartite information which is defined by (\ref{3par}) and in terms of the mutual information it is given by \eqref{3par1}.\footnote{ In the context of classical information theory, the tripartite information which is used as a measure of entanglement is also called the $I$-measure \cite{book}. }
	In \cite{Hayden:2011ag} it is shown that for quantum field theories
	with holographic duals, the HMI is  monogamous due to $I^{[3]}({A}:{B}:{C})\leq0$. For $n$ strips with the same width $\ell$ separated by distance $h$, in the specific limit where $h\ll\ell$, in \cite{Alishahiha:2014jxa} this property has been extended for the holographic $n$-partite information, namely the $n$-partite information which is defined by (\ref{npar2}) has a definite sign: it is positive (negative) for even (odd) $n$. In this section by relaxing the proposed limit for $\ell$ and $h$, we verify this property for mutual and tripartite information in AdS background and then we generalize the results to four and five partite information. To fix the notation, let us suppose a $d$ dimensional strip as an entangling region with width $\ell$ which is defined by
	\begin{equation}\label{strip}
		- \frac{\ell}{2} \le {x_1} \le
		\frac{\ell}{2},\,\,\,\,\,\,\,\,\,\,\,\,\,\,\, 0 \le {x_i} \le
		L\,\,\,\,\,\left( {i = 2,\cdots,d - 1}
		\right),\,\,\,\,\,\,\,\,\,\,\,\,\,\,\, \mbox{t = fixed}.
	\end{equation}
	It is noted that the strip entangling regions are considered to be symmetric meaning that they have the same width $\ell$ and are all separated by distance $h$. To describe the geometric background, let us use the following metric
	\begin{equation}\label{bulk}
		d{s^2} = \frac{1}{{{\rho ^2}}}\left( { - f(\rho )d{t^2} + \frac{{d{\rho ^2}}}{{f(\rho )}} + \sum\limits_{i = 1}^{d - 1} {d{x_i}^2} }
		\right)
	\end{equation}
	where one obtains an AdS geometry by setting $f(\rho ) = 1$ and AdS-BH with  $f(\rho ) = 1 - {\left( {\frac{\rho }{{{\rho_H}}}} \right)^d}$ in which ${\rho _H}$ is the horizon radius.\footnote{In this paper we set the AdS radius to one.} According to the AdS/CFT correspondence, the AdS and AdS-BH geometries respectively correspond to a vacuum and thermal states of a CFT. To study the holographic $n$-partite information, one can use (\ref{npar}) in which  $n$-partite information is given by HEE. Relegating the details of calculation to the Appendix, we present the final result of HEE in AdS, which is given by \cite{Alishahiha:2014jxa}
\begin{equation}
		{S_{\text{vac}}}\left( \ell \right)=
		\begin{cases}
			\frac{1}{{4{G_N}}}\ln \frac{\ell}{\epsilon}, & \quad d = 2  \\
			\frac{{{L^{d - 2}}}}{{4{G_N}}}\left( {\frac{1}{{(d - 2){\epsilon ^{d - 2}}}} - \frac{{{c_0}(d)}}{{{\ell^{d - 2}}}}} \right), & \quad d > 2  \\
		\end{cases}\label{svac}
	\end{equation}
in which $\epsilon$ denotes UV cut-off of the QFT and ${c_0}\left( d \right)$ is given by
\[{c_0}\left( d \right) = \frac{{{2^{d - 2}}}}{{d - 2}}{\left( {\sqrt \pi  \frac{{\Gamma \left( {\frac{d}{{2d - 2}}} \right)}}{{\Gamma \left( {\frac{1}{{2d - 2}}} \right)}}} \right)^{d - 1}}.\]
In the rest of this section we study the holographic $n$-partite information in AdS.  Note that in all of the following numeric computations, we assume that ${\rho _H} = 1$, $L=1$ and $1/4{G_N} = 1$, for simplicity.
\subsection{Holographic mutual and tripartite information}

\begin{figure}[h!]
	\centering
	\begin{tikzpicture}[scale=.55]
	\draw[ultra thick,c1] (0,0) -- (2,0);
	\draw[ultra thick,c1] (2.5,0) -- (4.5,0);
	\draw[ultra thick,aqua] (2,0) arc (0:180:1cm);
	\draw[ultra thick,aqua] (4.5,0) arc (0:180:1cm);
	\draw[ultra thick,c1] (10,0) -- (12,0);
	\draw[ultra thick,c1] (12.5,0) -- (14.5,0);
	\draw[ultra thick,aqua] (14.5,0) arc (0:180:2.25cm);
	\draw[ultra thick,aqua] (12.5,0) arc (0:180:0.25cm);
	\draw[] (-1,0.1) node[left] {${\cal A}_{1}$};
	\draw[] (9,0.1) node[left] {${\cal A}_{2}$};
	\draw[] (1,0) node[below] {$A$};
	\draw[] (3.5,0) node[below] {$B$};
	\draw[] (11,0) node[below] {$A$};
	\draw[] (13.5,0) node[below] {$B$};
	\end{tikzpicture}\caption{ A schematic diagram of the holographic prescription of two different configurations for computing $S(A\cup B)$.}\label{figure1}
\end{figure}
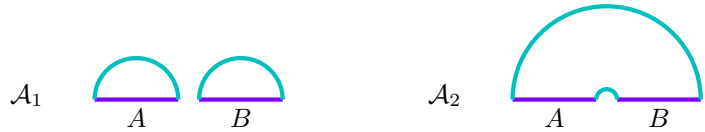

\begin{figure}[h]
		\centering
		\includegraphics[width=5.5cm]{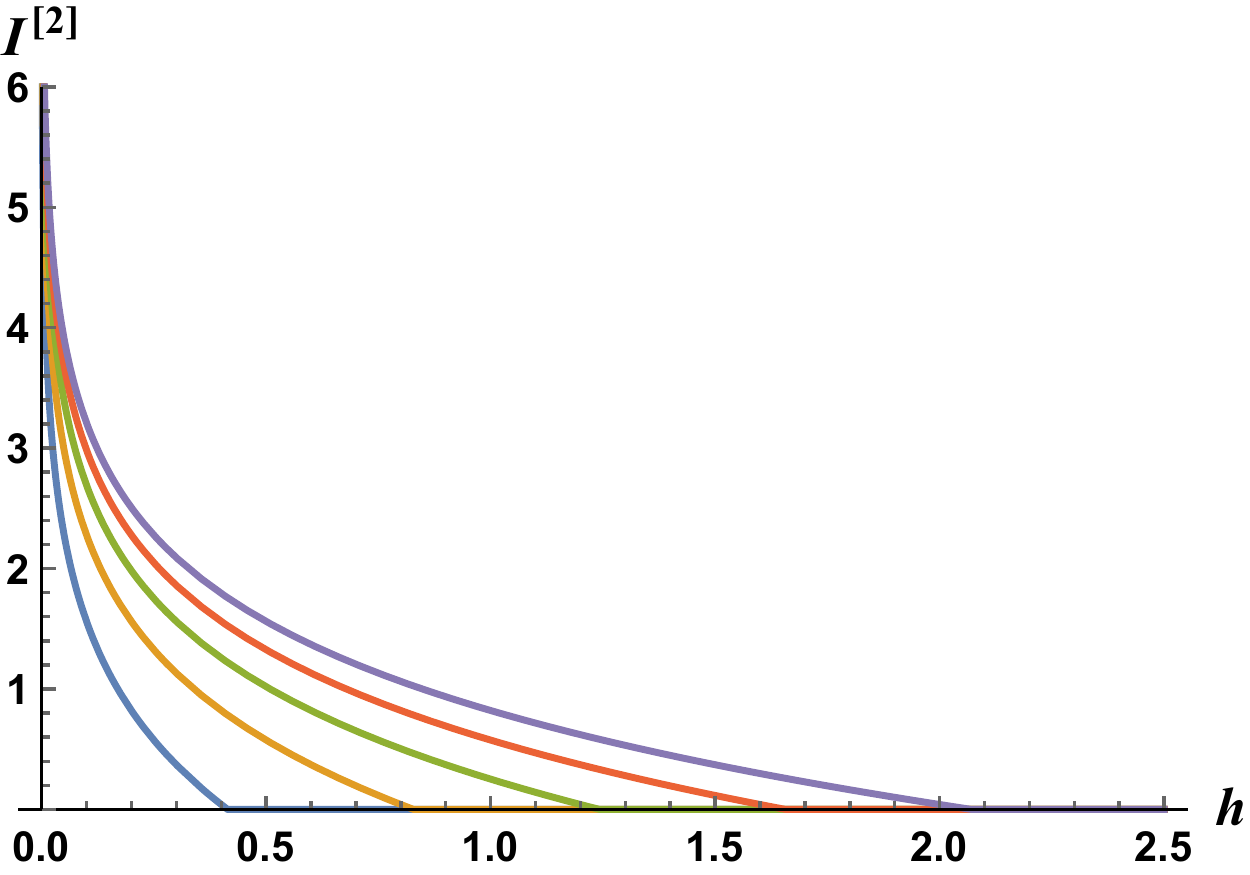}\,\,\,\,\,\,\,\includegraphics[width=5.5cm]{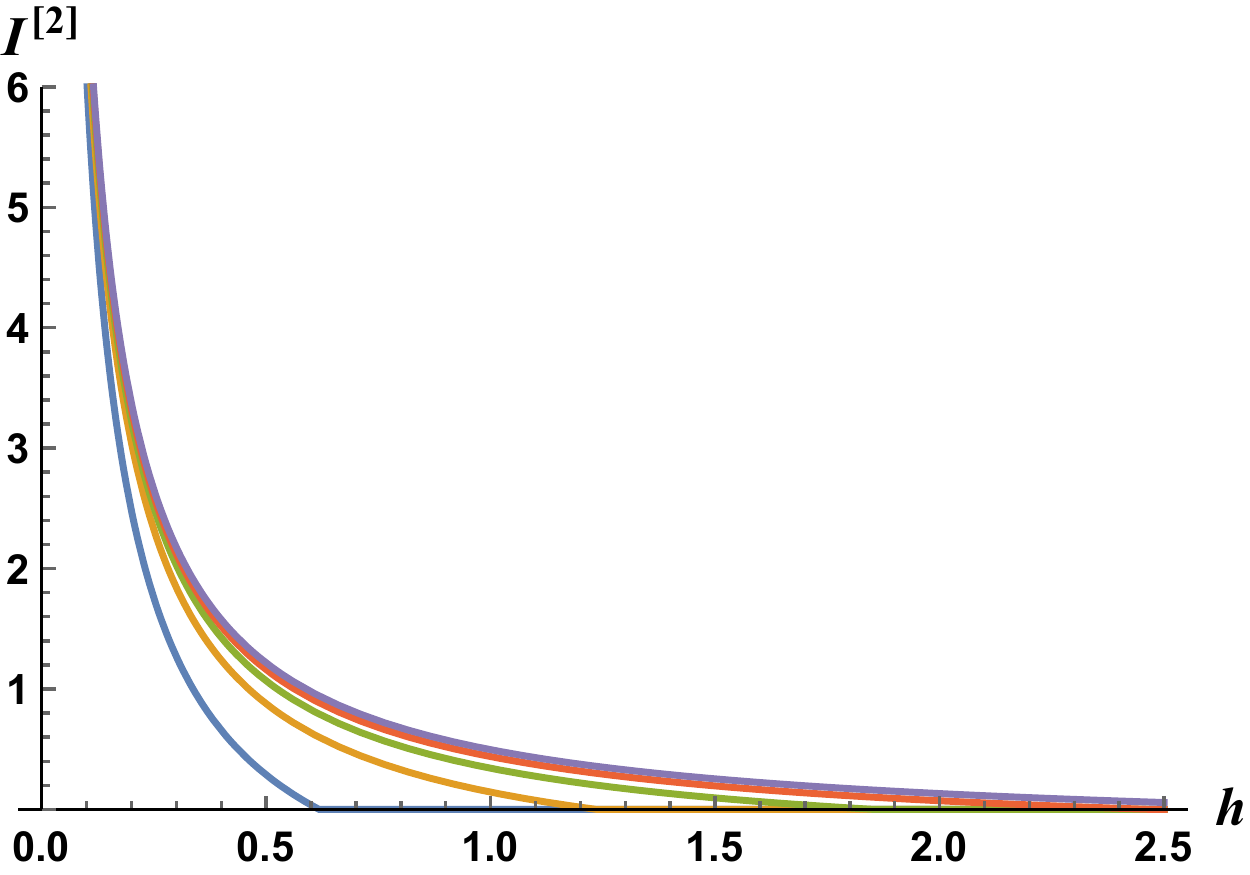}
		\caption{Holographic mutual information in ${\rm{Ad}}{{\rm{S}}_3}$ (\emph{left}) and ${\rm{Ad}}{{\rm{S}}_4}$ (\emph{right})
			backgrounds for $\ell = 1,{\rm{ }}2,{\rm{ }}3,{\rm{ }}4,{\rm{ }}5$ from left to right.}\label{n2ads}
	\end{figure}

Holographic mutual and tripartite information can be written in terms of
holographic entanglement entropies which are given by equations (\ref{HMI}) and (\ref{3par}). In the context of holographic CFTs, it was pointed out mutual information undergoes a first order phase transition due to a discontinuity in its first derivative \cite{Headrick:2010zt}. In fact for a given two subsystems $A$ and $B$,  when they are close to each other there is a finite correlation between them which means $I(A: B)\neq0$, however, as the separation between them is increased, the mutual correlation vanishes. In fact when $I(A: B) = 0$, the two sub-systems become completely decoupled, this is actually called disentangling transition \cite{Fischler:2012uv}. It is claimed that this phase transition is related to the
large central charge limit of the CFT and by considering quantum corrections it would disappear \cite{Faulkner:2013yia}. There is a simple gravity picture for this phase transition depending on a jump between two different configuration
candidates for the minimal surface of the entanglement entropy of the union region $S(A\cup B)$. Actually in our case for two strips, in the bulk there are always two candidates for minimal surfaces which are schematically shown in Fig.\ref{figure1} and depending on the separation between $A$ and $B$, one should use one of them in computing $S(A\cup B)$.\footnote{Note that there are other possibilities where the two minimal area surfaces cross each other, but one can show that such cross surfaces have indeed a larger area than those depicted in Fig.\ref{figure1}.}
Therefore depending on the ratio of $\frac{h}{\ell}$, one obtains
\begin{equation}\label{HEE-forms}
	S({A} \cup {B}) = \left\{ \begin{array}{l}
	 S\left( h \right) + S\left( {2\ell  + h} \right) (\equiv {{\cal A}_2})\,\,\,\,\,\,\text{for $\frac{h}{\ell}$ is smaller than the critical ratio},\\
		2S\left( \ell  \right) (\equiv {{\cal A}_1})\hspace{22mm}\text{for $\frac{h}{\ell}$ is larger than the critical ratio}, \\
	\end{array} \right.
\end{equation}
from which the minimum one is used in computing HMI. The critical ratio, $r_1$, at which the minimum configuration is transitioned from ${{\cal A}_2}$ to ${{\cal A}_1}$ can be obtained by equating ${\cal A}_1$ with ${\cal A}_2$, it is in fact a number that depends on the dimension of CFT. In AdS$_3$ and AdS$_4$, we have listed this specific value in Table.\ref{n2n3table}. Consequently, for two disjoint entangling regions, the holographic mutual information becomes
\begin{equation}\label{i2}
	I\left( {h,\ell} \right) = \left\{ {\begin{array}{*{20}{c}}
		{{2S}\left( \ell \right) - {S}\left( h \right) - {S}\left( {h + 2\ell} \right)}, \hfill & {\quad 0 < \frac{h}{\ell} < {r_1}}, \hfill \\
			0, \hfill & {\quad {r_1} \le \frac{h}{\ell}}. \hfill  \\
			\end{array}} \right.
	\end{equation}
 It can be shown that HMI is positive in ranges smaller than the critical
	value (denoted by $r_i$) and reaches zero in greater ranges; this is illustrated in Fig.\ref{n2ads}.
	
	About tripartite information more than two configurations are
	possible. In fact, for $n$ entangling regions with no presumptions, the
	overall number of configurations is $ \left( {2n - 1} \right)!!$;
	however, since the strip regions are assumed to be symmetric, some
	of the configurations would have equivalent areas so the number of
	possible configurations will be reduced and the rest can be
	obtained by rearranging these ones \cite{Allais:2011ys,
		Alishahiha:2014jxa}. For $n=3$, there are seven independent
	configurations, but one can show that
		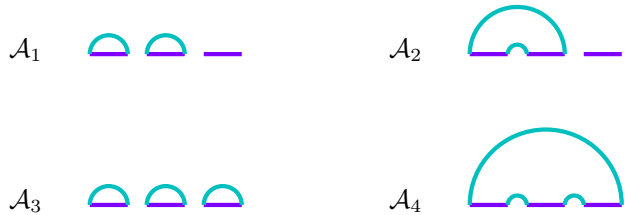
\begin{figure}[h!]
		\centering
		\begin{tikzpicture}[scale=.5]
		\draw[ultra thick,c1] (7,-4) -- (8,-4);
		\draw[ultra thick,c1] (8.5,-4) -- (9.5,-4);
		\draw[ultra thick,c1] (10,-4) -- (11,-4);
		\draw[ultra thick,aqua] (8,-4) arc (0:180:.5cm);
		\draw[ultra thick,aqua] (9.5,-4) arc (0:180:.5cm);
		\draw[] (6,-3.9) node[left] {${\cal A}_{1}$};
		
		\draw[ultra thick,c1] (17,-4) -- (18,-4);
		\draw[ultra thick,c1] (18.5,-4) -- (19.5,-4);
		\draw[ultra thick,c1] (20,-4) -- (21,-4);
		\draw[ultra thick,aqua] (18.5,-4) arc (0:180:.25cm);
		\draw[ultra thick,aqua] (19.5,-4) arc (0:180:1.25cm);
		\draw[] (16,-3.9) node[left] {${\cal A}_{2}$};
		
		\draw[ultra thick,c1] (7,-8) -- (8,-8);
		\draw[ultra thick,c1] (8.5,-8) -- (9.5,-8);
		\draw[ultra thick,c1] (10,-8) -- (11,-8);
		\draw[ultra thick,aqua] (8,-8) arc (0:180:.5cm);
		\draw[ultra thick,aqua] (9.5,-8) arc (0:180:.5cm);
		\draw[ultra thick,aqua] (11,-8) arc (0:180:.5cm);
		\draw[] (6,-7.9) node[left] {${\cal A}_{3}$};
		
		\draw[ultra thick,c1] (17,-8) -- (18,-8);
		\draw[ultra thick,c1] (18.5,-8) -- (19.5,-8);
		\draw[ultra thick,c1] (20,-8) -- (21,-8);
		\draw[ultra thick,aqua] (18.5,-8) arc (0:180:.25cm);
		\draw[ultra thick,aqua] (20,-8) arc (0:180:.25cm);
		\draw[ultra thick,aqua] (21,-8) arc (0:180:2cm);
		\draw[] (16,-7.9) node[left] {${\cal A}_{4}$};
		\end{tikzpicture}\caption{Schematic representation of RT surfaces corresponding to the possible configurations of minimal surfaces in computing holographic tripartite information.}\label{imutual}
	\end{figure}
	only four different configurations, which are schematically shown in Fig.\ref{imutual}, are competing in selecting the minimum area. In the present case, one obtains two critical values for the ratio of $h/\ell$ in which the minimal area turns from one to another configuration. Precisely depending on the separation between entangling regions, there are a competition between ${\cal A}_1$ and ${\cal A}_2$ and also between  ${\cal A}_3$ and ${\cal A}_4$. In Table.\ref{n2n3table},  we have listed these critical ratios in AdS$_3$ and AdS$_4$.
	\begin{table}[h!]
		\caption{Critical points for mutual information ($n=2$) and tripartite information ($n=3$) in AdS background.}
		\vspace{5pt}
		\centering
		\begin{tabular}{c|c|cc}
			\hline
			\head{\shortstack{\\Holographic information}} & \head{Background} & \head{${r_1}$} & \head{${r_2}$}\\[5pt]
			\hline
			\shortstack{\\Mutual information} & ${\rm{Ad}}{{\rm{S}}_3}$ & $- 1 + \sqrt 2$ &-\\[5pt]
			{\rm{ }} & ${\rm{Ad}}{{\rm{S}}_4}$ & $\frac{{ - 1 + \sqrt 5 }}{2}$&-\\[5pt]
			\hline
			\shortstack{\\Tripartite information} & ${\rm{Ad}}{{\rm{S}}_3}$ & $- 1 + \sqrt 2$ & 0.5\\[5pt]
			{\rm{ }} & ${\rm{Ad}}{{\rm{S}}_4}$ & $\frac{{ - 1 + \sqrt 5 }}{2}$ & $\frac{{ - 1 + \sqrt {10} }}{3}$\\[5pt]
			\hline
		\end{tabular}\label{n2n3table}
	\end{table}
	\\ Consequently, we have found the holographic tripartite information as follows
	\begin{equation}\label{i3}
		{I^{\left[ 3 \right]}}\left( {h,\ell} \right) = \left\{ {\begin{array}{*{20}{c}}
				{{S}\left( \ell \right) - 2{S}\left( {h + 2\ell} \right) + {S}\left( {2h + 3\ell} \right)}, \hfill & {\quad 0 < \frac{h}{\ell} < {r_1}} \hfill  \\
				{2{S}\left( h \right) - 3{S}\left( \ell \right) + {S}\left( {2h + 3\ell} \right)}, \hfill & {\quad {r_1} \le \frac{h}{\ell} < {r_2}} \hfill  \\
				0, \hfill & {\quad {r_2} \le \frac{h}{\ell}} \hfill  \\
			\end{array}} \right.
		\end{equation}
		For different values of $\ell$, Fig.\ref{n3ads} indicates that the tripartite information is always negative \begin{equation}
			I^{[3]}(A:B:C)\leq 0.
		\end{equation} This numeric result shows that according to the RT formula, the holographic tripartite information is always extensive which is in accordance with \cite{Hayden:2011ag}.
		\begin{figure}[h!]
			\centering
			\includegraphics[width=6.5cm]{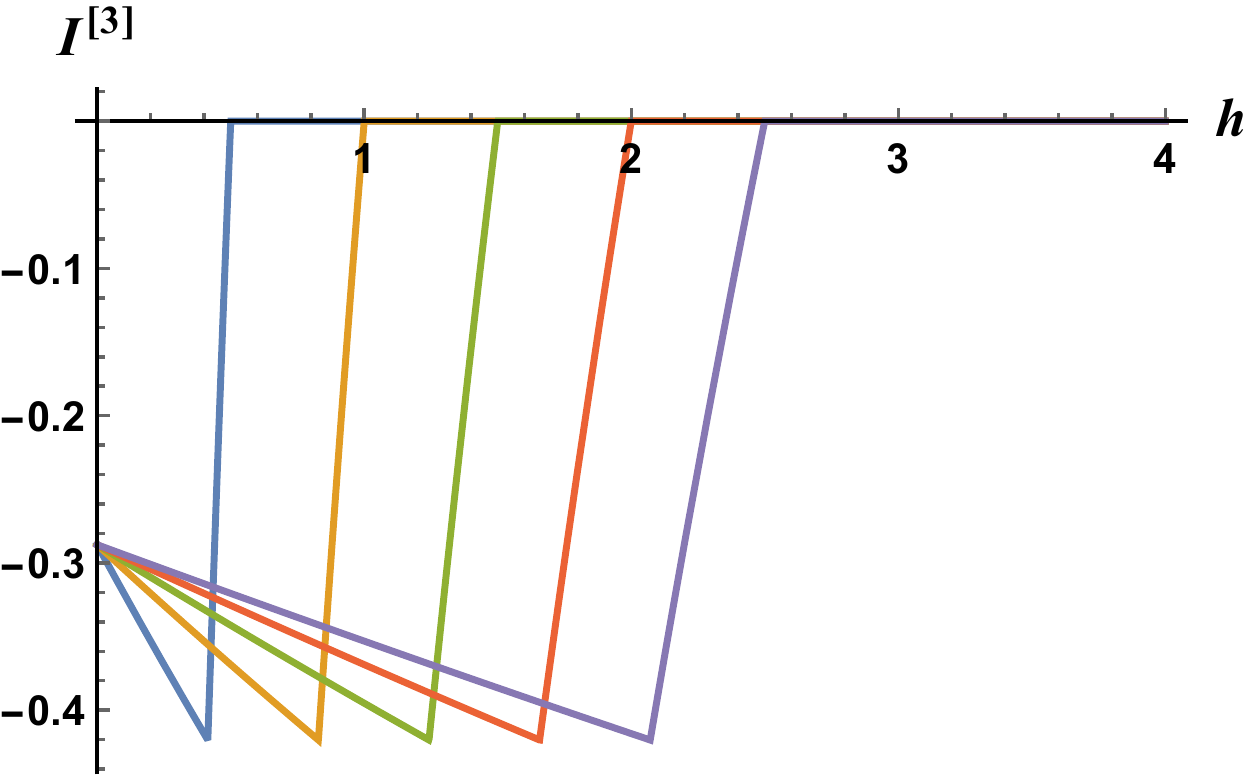}\,\,\,\,\,\,\,\includegraphics[width=6.5cm]{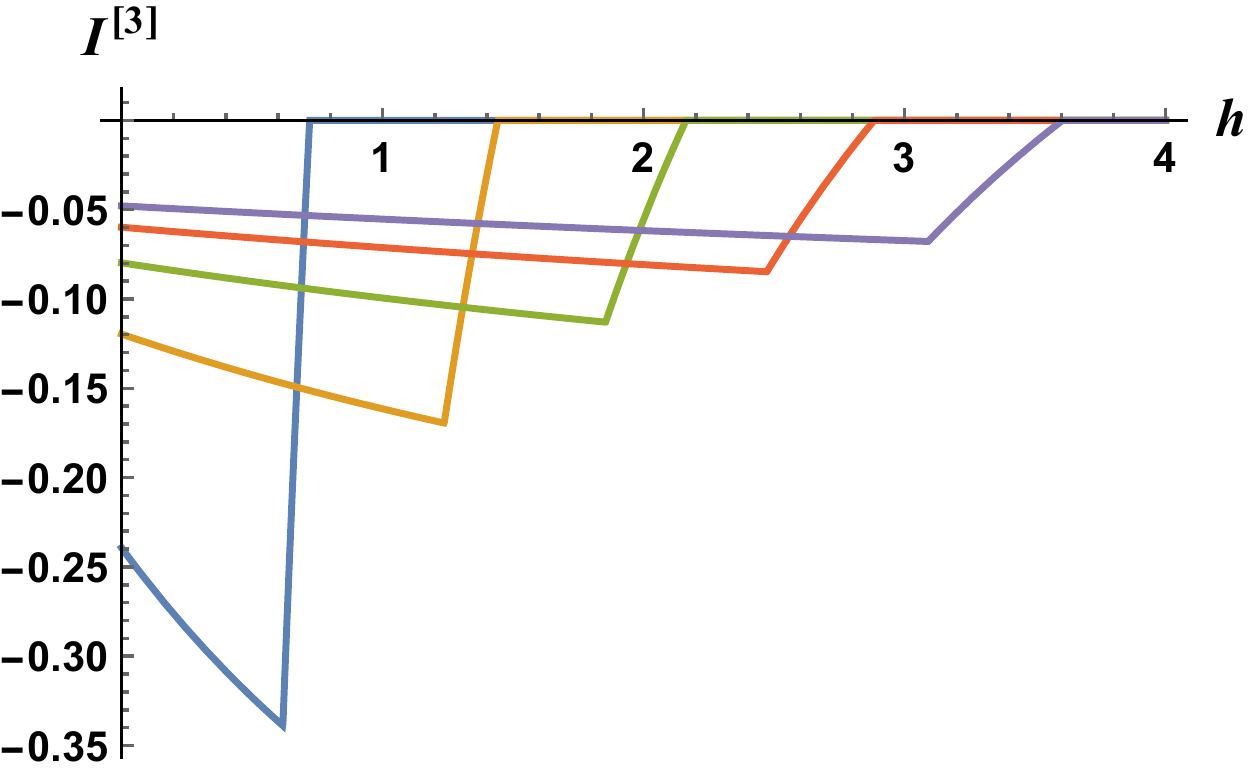}
			\caption{Holographic tripartite information in ${\rm{Ad}}{{\rm{S}}_3}$ (\emph{left}) and ${\rm{Ad}}{{\rm{S}}_4}$ (\emph{right})
				backgrounds for $\ell = 1,{\rm{ }}2,{\rm{ }}3,{\rm{ }}4,{\rm{ }}5$ from left to right.}\label{n3ads}
		\end{figure}
		

\subsection{Holographic 4-partite information}

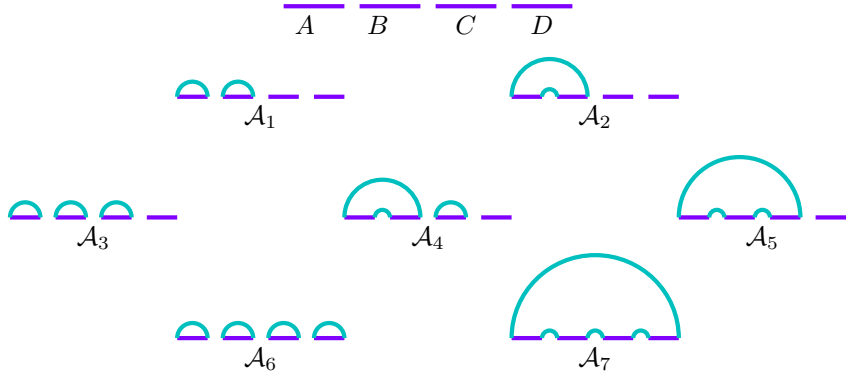
\begin{figure}[h!]
	\centering
	\begin{tikzpicture}[scale=.4]
	
	\draw[ultra thick,c1] (9,-9) -- (11,-9);
	\draw[ultra thick,c1] (11.5,-9) -- (13.5,-9);
	\draw[ultra thick,c1] (14,-9) -- (16,-9);
	\draw[ultra thick,c1] (16.5,-9) -- (18.5,-9);
	\draw[] (9.7,-9) node[below] {$A$};
	\draw[] (12.1,-9) node[below] {$B$};
	\draw[] (15,-9) node[below] {$C$};
	\draw[] (17.5,-9) node[below] {$D$};
	\draw[ultra thick,c1] (5.5,-12) -- (6.5,-12);
	\draw[ultra thick,c1] (7,-12) -- (8,-12);
	\draw[ultra thick,c1] (8.5,-12) -- (9.5,-12);
	\draw[ultra thick,c1] (10,-12) -- (11,-12);
	\draw[ultra thick,aqua] (6.5,-12) arc (0:180:0.5cm);
	\draw[ultra thick,aqua] (8,-12) arc (0:180:.5cm);
	\draw[] (8.25,-12) node[below] {${\cal A}_{1}$};
	
	\draw[ultra thick,c1] (16.5,-12) -- (17.5,-12);
	\draw[ultra thick,c1] (18,-12) -- (19,-12);
	\draw[ultra thick,c1] (19.5,-12) -- (20.5,-12);
	\draw[ultra thick,c1] (21,-12) -- (22,-12);
	\draw[ultra thick,aqua] (18,-12) arc (0:180:0.25cm);
	\draw[ultra thick,aqua] (19,-12) arc (0:180:1.25cm);
	\draw[] (19.25,-12) node[below] {${\cal A}_{2}$};

	\draw[ultra thick,c1] (0,-16) -- (1,-16);
	\draw[ultra thick,c1] (1.5,-16) -- (2.5,-16);
	\draw[ultra thick,c1] (3,-16) -- (4,-16);
	\draw[ultra thick,c1] (4.5,-16) -- (5.5,-16);
	\draw[ultra thick,aqua] (1,-16) arc (0:180:.5cm);
	\draw[ultra thick,aqua] (2.5,-16) arc (0:180:.5cm);
	\draw[ultra thick,aqua] (4,-16) arc (0:180:.5cm);
	\draw[] (2.75,-16) node[below] {${\cal A}_{3}$};
	
	\draw[ultra thick,c1] (11,-16) -- (12,-16);
	\draw[ultra thick,c1] (12.5,-16) -- (13.5,-16);
	\draw[ultra thick,c1] (14,-16) -- (15,-16);
	\draw[ultra thick,c1] (15.5,-16) -- (16.5,-16);
	\draw[ultra thick,aqua] (15,-16) arc (0:180:.5cm);
	\draw[ultra thick,aqua] (12.5,-16) arc (0:180:.25cm);
	\draw[ultra thick,aqua] (13.5,-16) arc (0:180:1.25cm);
	\draw[] (13.75,-16) node[below] {${\cal A}_{4}$};
	
	\draw[ultra thick,c1] (22,-16) -- (23,-16);
	\draw[ultra thick,c1] (23.5,-16) -- (24.5,-16);
	\draw[ultra thick,c1] (25,-16) -- (26,-16);
	\draw[ultra thick,c1] (26.5,-16) -- (27.5,-16);
	\draw[ultra thick,aqua] (23.5,-16) arc (0:180:.25cm);
	\draw[ultra thick,aqua] (25,-16) arc (0:180:.25cm);
	\draw[ultra thick,aqua] (26,-16) arc (0:180:2cm);
	\draw[] (24.75,-16) node[below] {${\cal A}_{5}$};

	\draw[ultra thick,c1] (5.5,-20) -- (6.5,-20);
	\draw[ultra thick,c1] (7,-20) -- (8,-20);
	\draw[ultra thick,c1] (8.5,-20) -- (9.5,-20);
	\draw[ultra thick,c1] (10,-20) -- (11,-20);
	\draw[ultra thick,aqua] (6.5,-20) arc (0:180:.5cm);
	\draw[ultra thick,aqua] (8,-20) arc (0:180:.5cm);
	\draw[ultra thick,aqua] (9.5,-20) arc (0:180:.5cm);
	\draw[ultra thick,aqua] (11,-20) arc (0:180:.5cm);
	\draw[] (8.25,-20) node[below] {${\cal A}_{6}$};
	
	\draw[ultra thick,c1] (16.5,-20) -- (17.5,-20);
	\draw[ultra thick,c1] (18,-20) -- (19,-20);
	\draw[ultra thick,c1] (19.5,-20) -- (20.5,-20);
	\draw[ultra thick,c1] (21,-20) -- (22,-20);
	\draw[ultra thick,aqua] (18,-20) arc (0:180:.25cm);
	\draw[ultra thick,aqua] (19.5,-20) arc (0:180:.25cm);
	\draw[ultra thick,aqua] (21,-20) arc (0:180:.25cm);
	\draw[ultra thick,aqua] (22,-20) arc (0:180:2.75cm);
	\draw[] (19.25,-20) node[below] {${\cal A}_{7}$};
	\end{tikzpicture}\caption{Schematic representation of the RT surfaces corresponding to seven different configurations in computing the HEE of union four strips.}\label{fourpar}
\end{figure}

In this subsection we consider the holographic 4-partite information
in AdS background. To do so, we follow the same approach of previous subsection and hence the first step would be considering all of the possible configurations for the unions of
entangling regions. In fact, for four strips $A,B,C$ and $D$ as the entangling regions with the same width $\ell$ separated by a distance $h$, the four partite information is given by
\begin{eqnarray}
I^{[4]}(h,\ell)&=&4S(A)-3S(A\cup B)-2S(A\cup C)-S(A\cup D)+2S({A\cup B\cup C})+2S({A\cup B\cup C})\nonumber\\
&-&S({A \cup B\cup C\cup D}).
\end{eqnarray} For this case there are 19 independent
configurations \cite{Alishahiha:2014jxa} from which choosing those that produce minimum area is
the next and most important stage. In the bulk for computing the related HEE, there are seven candidates for minimal surfaces which are schematically shown in Fig.\ref{fourpar}.

\begin{figure}[h!]
		\centering
		\includegraphics[width=6.5cm]{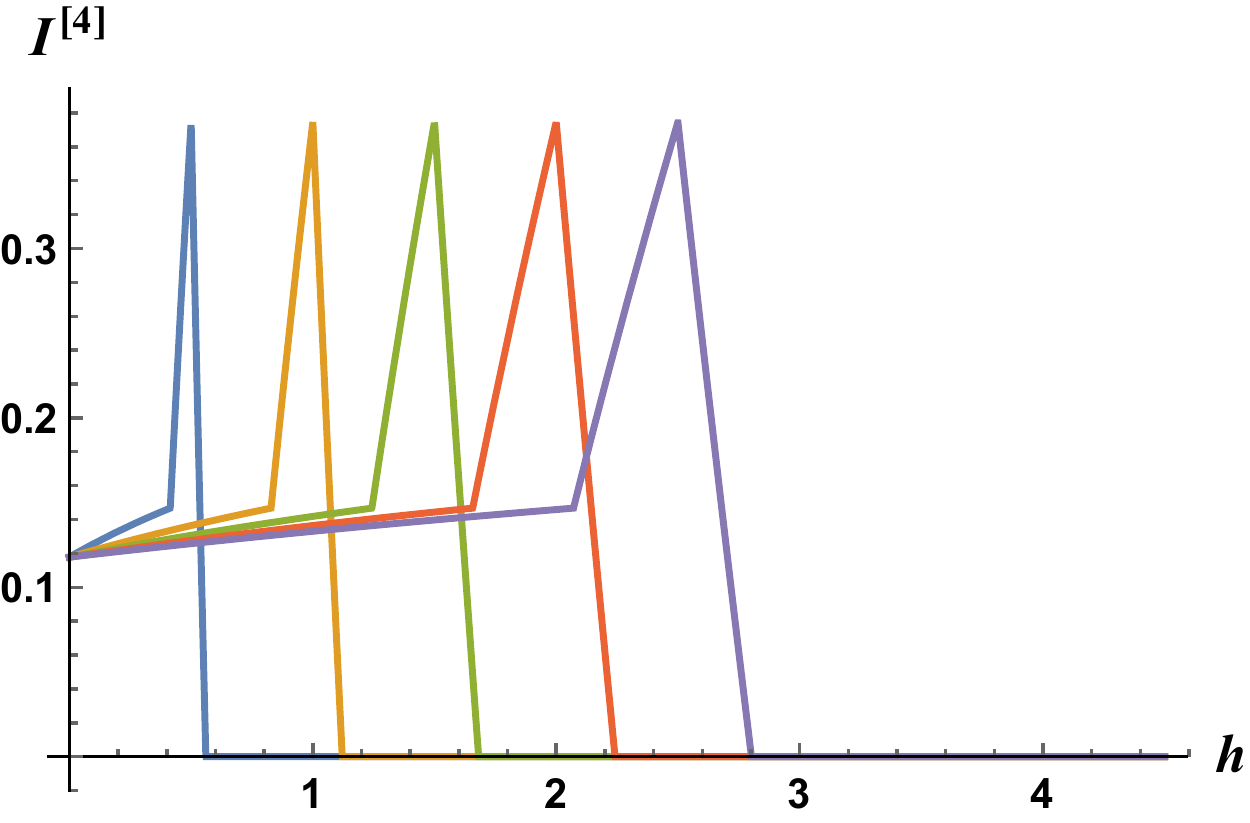}\,\,\,\,\,\,\,\includegraphics[width=6.5cm]{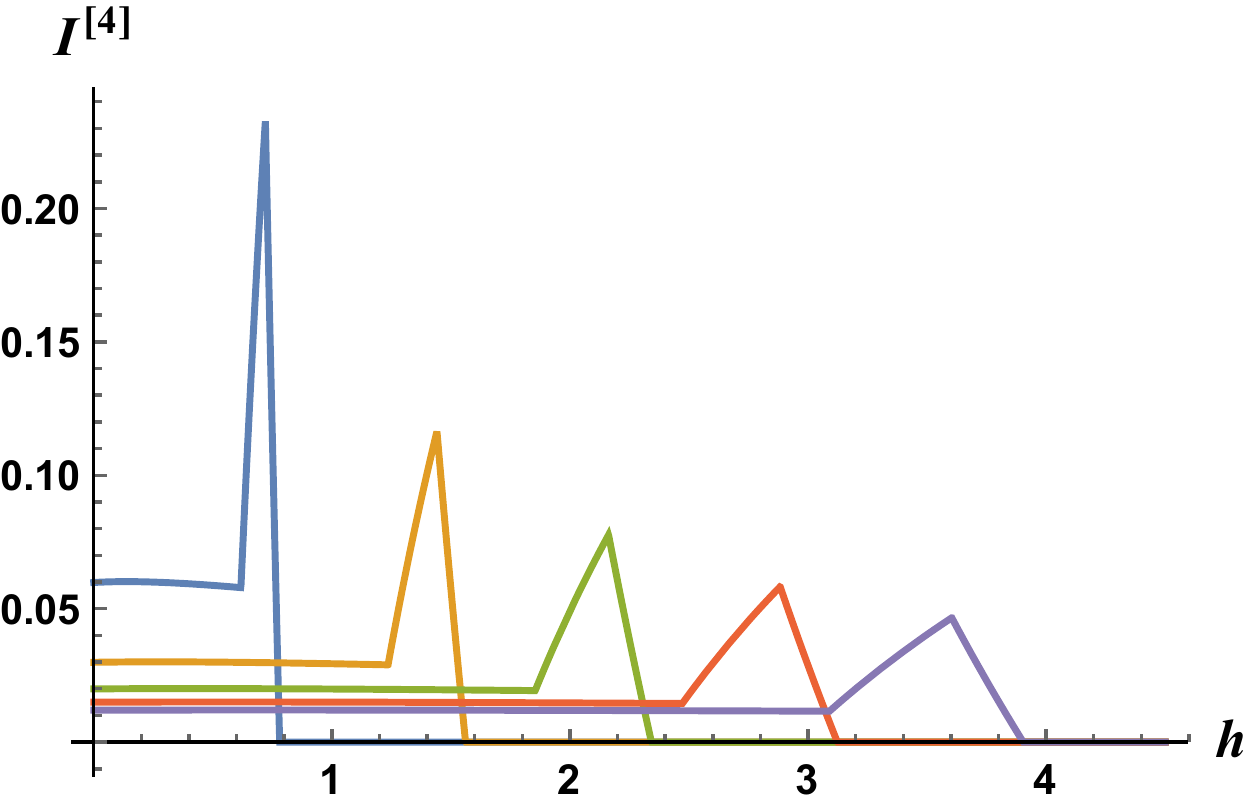}
		\caption{Holographic 4-partite information in ${\rm{Ad}}{{\rm{S}}_3}$ (\emph{left}) and ${\rm{Ad}}{{\rm{S}}_4}$
			(\emph{right}) backgrounds for $\ell = 1,{\rm{ }}2,{\rm{ }}3,{\rm{ }}4,{\rm{ }}5$ from left to right.}\label{n4ads}
	\end{figure}

In the computation of related minimal surfaces, one finds three critical values in  AdS$_3$ and AdS$_4$
backgrounds which are listed in Table.\ref{n4table}. The first one
is caused by transitions from ${\cal A}_{4}$ to ${\cal A}_{3}$ and also ${\cal A}_{2}$ to
${\cal A}_{1}$. The second critical point is related to the transition of minimal area between ${\cal A}_{5}$ and ${\cal A}_{3}$ and finally the transition between ${\cal A}_{7}$ and ${\cal A}_6$ gives us the third one.
\begin{table}[h!]
	\caption{Critical points for $n=4$ in AdS background.}
	\vspace{5pt}
	\centering
	\begin{tabular}{c|ccc}
		\hline
		\head{\shortstack{\\Geometric background}} & \head{${r_1}$} & \head{${r_2}$} & \head{${r_3}$}\\[5pt]
		\hline
		\shortstack{\\${\rm{Ad}}{{\rm{S}}_3}$} & $- 1 + \sqrt 2$ & 0.5 & 0.560426 \\[5pt]
		${\rm{Ad}}{{\rm{S}}_4}$ & $\frac{{ - 1 + \sqrt 5 }}{2}$ & $\frac{{ - 1 + \sqrt {10} }}{3}$ & $\frac{{ - 1 + \sqrt {17} }}{4}$\\[5pt]
		\hline
	\end{tabular}\label{n4table}
\end{table}
Taking into consideration the minimal surface in each case and after making use of (\ref{npar}), one obtains the 4-partite information in the
resultant intervals as follows
\begin{equation}
	{I^{\left[ 4 \right]}}\left( {h,\ell} \right) = \left\{ {\begin{array}{*{20}{c}}
			{ {\rm{2}}S\left( {{\rm{2}}h{\rm{ + 3}}\ell} \right) - S\left( {h{\rm{ + 2}}\ell} \right) -
				S\left( {{\rm{3}}h{\rm{ + 4}}\ell} \right),} \hfill & {\quad 0 < \frac{h}{\ell} < {r_1}} \hfill  \\
			{S\left( h \right) - 2S\left( \ell \right) + 2S\left( {2h + 3\ell} \right) - S\left( {3h + 4\ell} \right),} \hfill & {\quad {r_1} \le \frac{h}{\ell} < {r_2}} \hfill  \\
			{ 4S\left( \ell \right) - 3S\left( h \right) - S\left( {3h + 4\ell} \right),} \hfill & {\quad {r_2} \le \frac{h}{\ell} < {r_3}} \hfill  \\
			{0,} \hfill & {\quad {r_3} \le \frac{h}{\ell}} \hfill  \\
		\end{array}}. \right.\label{i4}
	\end{equation}
	Now the main aim is to study the 4-partite information. Fig.\ref{n4ads} shows the numeric computation of 4-partite information for some fixed values of $\ell$ in AdS$_3$ and AdS$_4$ backgrounds which indicates that holographic 4-partite information is always positive.
	
\subsection{Holographic 5-partite information}

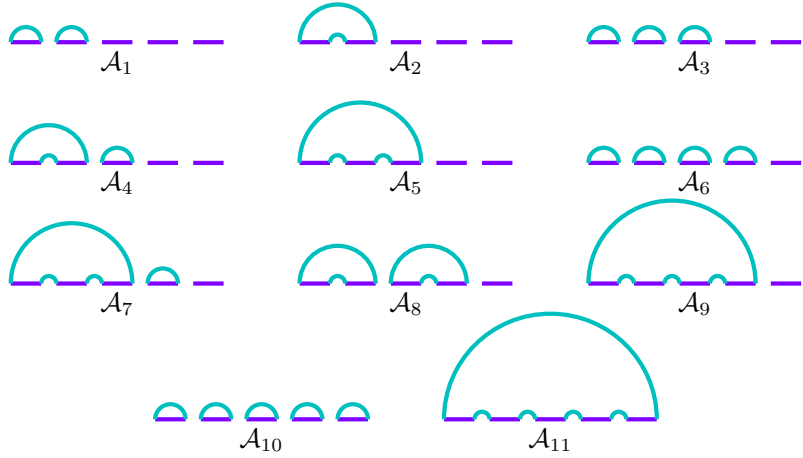
\begin{figure}[h!]
	\centering
	\begin{tikzpicture}[scale=.4]
	\draw[ultra thick,c1] (0,-24) -- (1,-24);
	\draw[ultra thick,c1] (1.5,-24) -- (2.5,-24);
	\draw[ultra thick,c1] (3,-24) -- (4,-24);
	\draw[ultra thick,c1] (4.5,-24) -- (5.5,-24);
	\draw[ultra thick,c1] (6,-24) -- (7,-24);
	\draw[ultra thick,aqua] (1,-24) arc (0:180:.5cm);
	\draw[ultra thick,aqua] (2.5,-24) arc (0:180:.5cm);
	\draw[] (3.5,-24) node[below] {${\cal A}_{1}$};
	
	\draw[ultra thick,c1] (9.5,-24) -- (10.5,-24);
	\draw[ultra thick,c1] (11,-24) -- (12,-24);
	\draw[ultra thick,c1] (12.5,-24) -- (13.5,-24);
	\draw[ultra thick,c1] (14,-24) -- (15,-24);
	\draw[ultra thick,c1] (15.5,-24) -- (16.5,-24);
	\draw[ultra thick,aqua] (11,-24) arc (0:180:.25cm);
	\draw[ultra thick,aqua] (12,-24) arc (0:180:1.25cm);
	\draw[] (13,-24) node[below] {${\cal A}_{2}$};
	
	\draw[ultra thick,c1] (19,-24) -- (20,-24);
	\draw[ultra thick,c1] (20.5,-24) -- (21.5,-24);
	\draw[ultra thick,c1] (22,-24) -- (23,-24);
	\draw[ultra thick,c1] (23.5,-24) -- (24.5,-24);
	\draw[ultra thick,c1] (25,-24) -- (26,-24);
	\draw[ultra thick,aqua] (20,-24) arc (0:180:.5cm);
	\draw[ultra thick,aqua] (21.5,-24) arc (0:180:.5cm);
	\draw[ultra thick,aqua] (23,-24) arc (0:180:.5cm);
	\draw[] (22.5,-24) node[below] {${\cal A}_{3}$};

	\draw[ultra thick,c1] (0,-28) -- (1,-28);
	\draw[ultra thick,c1] (1.5,-28) -- (2.5,-28);
	\draw[ultra thick,c1] (3,-28) -- (4,-28);
	\draw[ultra thick,c1] (4.5,-28) -- (5.5,-28);
	\draw[ultra thick,c1] (6,-28) -- (7,-28);
	\draw[ultra thick,aqua] (4,-28) arc (0:180:.5cm);
	\draw[ultra thick,aqua] (2.5,-28) arc (0:180:1.25cm);
	\draw[ultra thick,aqua] (1.5,-28) arc (0:180:.25cm);
	\draw[] (3.5,-28) node[below] {${\cal A}_{4}$};
	
	\draw[ultra thick,c1] (9.5,-28) -- (10.5,-28);
	\draw[ultra thick,c1] (11,-28) -- (12,-28);
	\draw[ultra thick,c1] (12.5,-28) -- (13.5,-28);
	\draw[ultra thick,c1] (14,-28) -- (15,-28);
	\draw[ultra thick,c1] (15.5,-28) -- (16.5,-28);
	\draw[ultra thick,aqua] (11,-28) arc (0:180:.25cm);
	\draw[ultra thick,aqua] (12.5,-28) arc (0:180:.25cm);
	\draw[ultra thick,aqua] (13.5,-28) arc (0:180:2cm);
	\draw[] (13,-28) node[below] {${\cal A}_{5}$};
	
	\draw[ultra thick,c1] (19,-28) -- (20,-28);
	\draw[ultra thick,c1] (20.5,-28) -- (21.5,-28);
	\draw[ultra thick,c1] (22,-28) -- (23,-28);
	\draw[ultra thick,c1] (23.5,-28) -- (24.5,-28);
	\draw[ultra thick,c1] (25,-28) -- (26,-28);
	\draw[ultra thick,aqua] (20,-28) arc (0:180:.5cm);
	\draw[ultra thick,aqua] (21.5,-28) arc (0:180:.5cm);
	\draw[ultra thick,aqua] (23,-28) arc (0:180:.5cm);
	\draw[ultra thick,aqua] (24.5,-28) arc (0:180:.5cm);
	\draw[] (22.5,-28) node[below] {${\cal A}_{6}$};

	\draw[ultra thick,c1] (0,-32) -- (1,-32);
	\draw[ultra thick,c1] (1.5,-32) -- (2.5,-32);
	\draw[ultra thick,c1] (3,-32) -- (4,-32);
	\draw[ultra thick,c1] (4.5,-32) -- (5.5,-32);
	\draw[ultra thick,c1] (6,-32) -- (7,-32);
	\draw[ultra thick,aqua] (1.5,-32) arc (0:180:.25cm);
	\draw[ultra thick,aqua] (3,-32) arc (0:180:.25cm);
	\draw[ultra thick,aqua] (4,-32) arc (0:180:2cm);
	\draw[ultra thick,aqua] (5.5,-32) arc (0:180:.5cm);
	\draw[] (3.5,-32) node[below] {${\cal A}_{7}$};
	
	\draw[ultra thick,c1] (9.5,-32) -- (10.5,-32);
	\draw[ultra thick,c1] (11,-32) -- (12,-32);
	\draw[ultra thick,c1] (12.5,-32) -- (13.5,-32);
	\draw[ultra thick,c1] (14,-32) -- (15,-32);
	\draw[ultra thick,c1] (15.5,-32) -- (16.5,-32);
	\draw[ultra thick,aqua] (11,-32) arc (0:180:.25cm);
	\draw[ultra thick,aqua] (12,-32) arc (0:180:1.25cm);
	\draw[ultra thick,aqua] (14,-32) arc (0:180:.25cm);
	\draw[ultra thick,aqua] (15,-32) arc (0:180:1.25cm);
	\draw[] (13,-32) node[below] {${\cal A}_{8}$};
	
	\draw[ultra thick,c1] (19,-32) -- (20,-32);
	\draw[ultra thick,c1] (20.5,-32) -- (21.5,-32);
	\draw[ultra thick,c1] (22,-32) -- (23,-32);
	\draw[ultra thick,c1] (23.5,-32) -- (24.5,-32);
	\draw[ultra thick,c1] (25,-32) -- (26,-32);
	\draw[ultra thick,aqua] (20.5,-32) arc (0:180:.25cm);
	\draw[ultra thick,aqua] (22,-32) arc (0:180:.25cm);
	\draw[ultra thick,aqua] (23.5,-32) arc (0:180:.25cm);
	\draw[ultra thick,aqua] (24.5,-32) arc (0:180:2.75cm);
	\draw[] (22.5,-32) node[below] {${\cal A}_{9}$};

	\draw[ultra thick,c1] (4.75,-36.5) -- (5.75,-36.5);
	\draw[ultra thick,c1] (6.25,-36.5) -- (7.25,-36.5);
	\draw[ultra thick,c1] (7.75,-36.5) -- (8.75,-36.5);
	\draw[ultra thick,c1] (9.25,-36.5) -- (10.25,-36.5);
	\draw[ultra thick,c1] (10.75,-36.5) -- (11.75,-36.5);
	\draw[ultra thick,aqua] (5.75,-36.5) arc (0:180:.5cm);
	\draw[ultra thick,aqua] (7.25,-36.5) arc (0:180:.5cm);
	\draw[ultra thick,aqua] (8.75,-36.5) arc (0:180:.5cm);
	\draw[ultra thick,aqua] (10.25,-36.5) arc (0:180:.5cm);
	\draw[ultra thick,aqua] (11.75,-36.5) arc (0:180:.5cm);
	\draw[] (8.25,-36.5) node[below] {${\cal A}_{10}$};
	
	\draw[ultra thick,c1] (14.25,-36.5) -- (15.25,-36.5);
	\draw[ultra thick,c1] (15.75,-36.5) -- (16.75,-36.5);
	\draw[ultra thick,c1] (17.25,-36.5) -- (18.25,-36.5);
	\draw[ultra thick,c1] (18.75,-36.5) -- (19.75,-36.5);
	\draw[ultra thick,c1] (20.25,-36.5) -- (21.25,-36.5);
	\draw[ultra thick,aqua] (15.75,-36.5) arc (0:180:.25cm);
	\draw[ultra thick,aqua] (17.25,-36.5) arc (0:180:.25cm);
	\draw[ultra thick,aqua] (18.75,-36.5) arc (0:180:.25cm);
	\draw[ultra thick,aqua] (20.25,-36.5) arc (0:180:.25cm);
	\draw[ultra thick,aqua] (21.25,-36.5) arc (0:180:3.5cm);
	\draw[] (17.75,-36.5) node[below] {${\cal A}_{11}$};	
	\end{tikzpicture}\caption{Schematic representation of the RT surfaces corresponding to eleven different configurations in computing the HEE of union five strips.}\label{fivepar}
\end{figure}

\begin{figure}[h!]
		\centering
		\includegraphics[width=6.5cm]{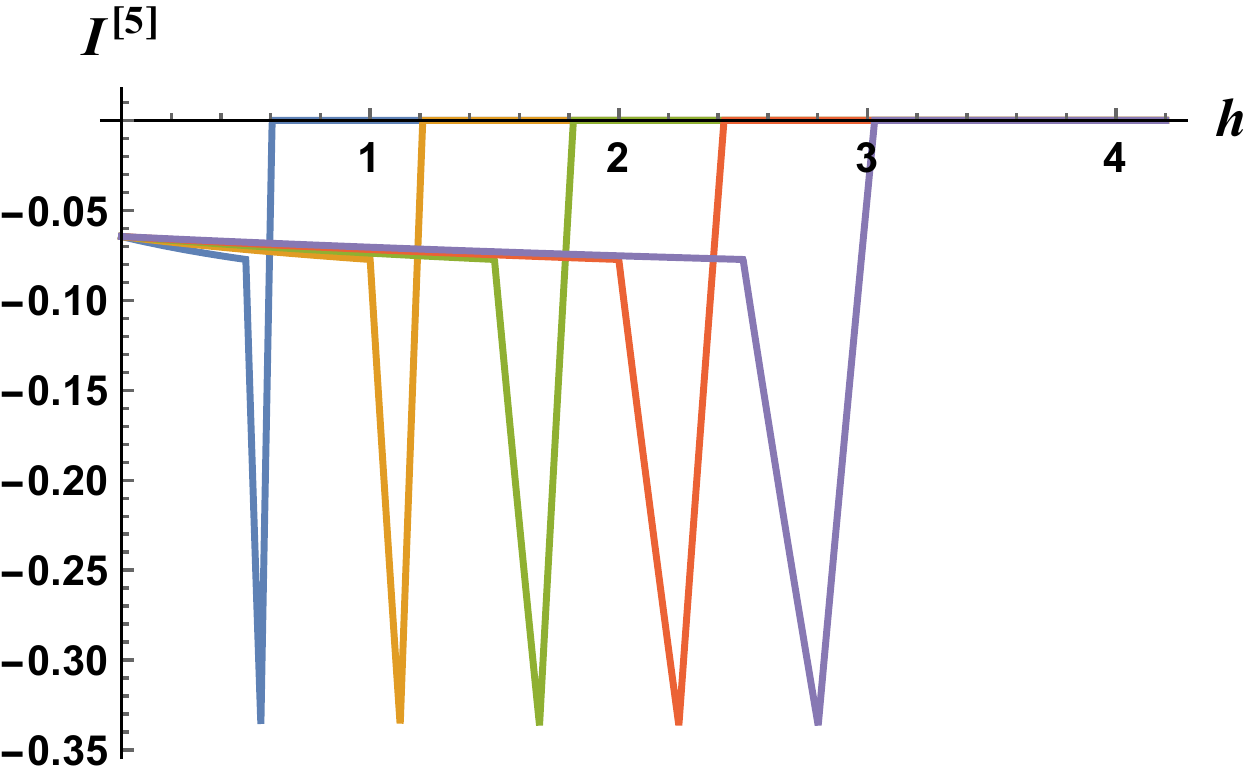}
		\caption{Holographic 5-partite information in ${\rm{Ad}}{{\rm{S}}_3}$ background for $\ell = 1,{\rm{ }}2,{\rm{ }}3,{\rm{ }}4,{\rm{ }}5$ from left to right.}\label{n5ads}
	\end{figure}

The holographic 5-partite information can indeed be defined by setting
$n=5$ in (\ref{npar}) and one can show that there are actually 50 independent configurations, however, among them only 11 configurations play the crucial role in computing the holographic 5-partite information in AdS$_3$ background as shown in Fig.\ref{fivepar}.
\\
One can show that in AdS$_3$ background, there are in fact four critical points which are resulted from the transitions
between different pairs of configurations as listed below:
\begin{itemize}
	\item $r_1$: ${\cal A}_{8}$ to ${\cal A}_{6}$,\,\,\, ${\cal A}_{4}$ to ${\cal A}_{3}$ \,\,\,and  ${\cal A}_{2}$ to ${\cal A}_{1}$,
	\item $r_2$: ${\cal A}_{7}$ to ${\cal A}_{6}$ and ${\cal A}_{5}$ to ${\cal A}_{3}$,
	\item $r_3$: ${\cal A}_{9}$ to ${\cal A}_{6}$,
	\item $r_4$: ${\cal A}_{11}$ to ${\cal A}_{10}$,
\end{itemize}
the critical points are summarized in Table.\ref{n5table}. Consequently, one can show that depending on the ratio of $\frac{h}{\ell}$, the holographic 5-partite information in AdS$_3$ background becomes
\begin{equation}
	{I^{\left[ 5 \right]}}\left( {h,\ell} \right) = \left\{ {\begin{array}{*{20}{c}}
			{S\left( {2h{\rm{ + 3}}\ell} \right) - {\rm{2}}S\left( {{\rm{3}}h{\rm{ + 4}}\ell} \right) + S\left( {4h{\rm{ + 5}}\ell} \right),} \hfill & {0 < \frac{h}{\ell} < {r_1}} \hfill  \\
			{S\left( {2h + 3\ell} \right) - 2S\left( {3h + 4\ell} \right) + S\left( {4h + 5\ell} \right),} \hfill & {{r_1} \le \frac{h}{\ell} < {r_2}} \hfill  \\
			{ 3S\left( \ell \right) - 2S\left( h \right) - 2S\left( {3h + 4\ell} \right) + S\left( {4h + 5\ell} \right),} \hfill & {{r_2} \le \frac{h}{\ell} < {r_3}} \hfill  \\
			{4S\left( h \right) - 5S\left( \ell \right) + S\left( {4h + 5\ell} \right),} \hfill & {{r_3} \le \frac{h}{\ell} < {r_4}} \hfill  \\
			{0,} \hfill & {{r_4} \le \frac{h}{\ell}} \hfill  \\
		\end{array}} \right. \label{i5ads3}
	\end{equation}
For some values of $\ell$,  Fig.\ref{n5ads} shows that holographic 5-partite information is always negative in AdS$_3$, namely
\begin{equation*}
I^{[5]}(A:B:C:D:E)\leq 0.
\end{equation*}
Making use of the above relation and \eqref{npar2}, one can write  
\begin{equation}
I^{[4]}(A:B:C:\,D\cup E)\geq I^{[4]}(A:B:C:\,D)+I^{[4]}(A:B:C:\,E),
\end{equation}
which indicates that holographic 4-partite information among five disjoint spatial regions $A$, $B$, $C$, $D$ and $E$ satisfies the inequality like monogamy relation.   
	\begin{table}[h!]
		\caption{Critical points for $n=5$ in AdS$_3$ background.}
		\vspace{5pt}
		\centering
		\begin{tabular}{c|ccccc}
			\hline
			\head{\shortstack{\\ Geometric background}} & \head{${r_1}$} & \head{${r_2}$} & \head{${r_3}$} & \head{${r_4}$}\\[5pt]
			\hline\\
			${\rm{Ad}}{{\rm{S}}_3}$ & $- 1 + \sqrt 2$ & 0.5 & 0.560426 & 0.60583\\[5pt]
			\hline
		\end{tabular}\label{n5table}
	\end{table}
	
	\section{Holographic $n$-partite information in AdS-BH geometry}
	
\begin{figure}[h!]
		\centering
		\includegraphics[width=6.5cm]{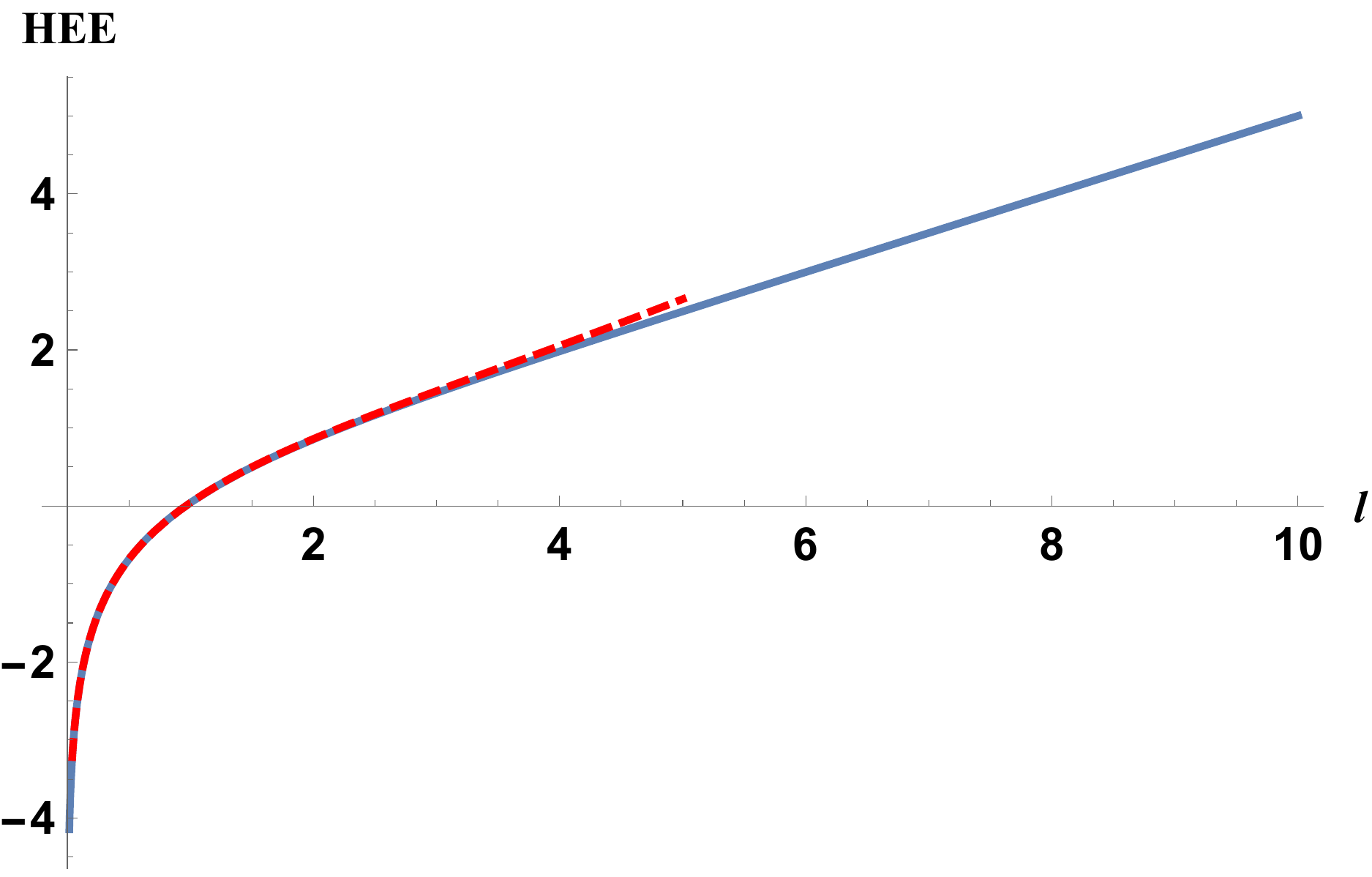}\,\,\,\,\,\,\,\,\,\,\includegraphics[width=6.5cm]{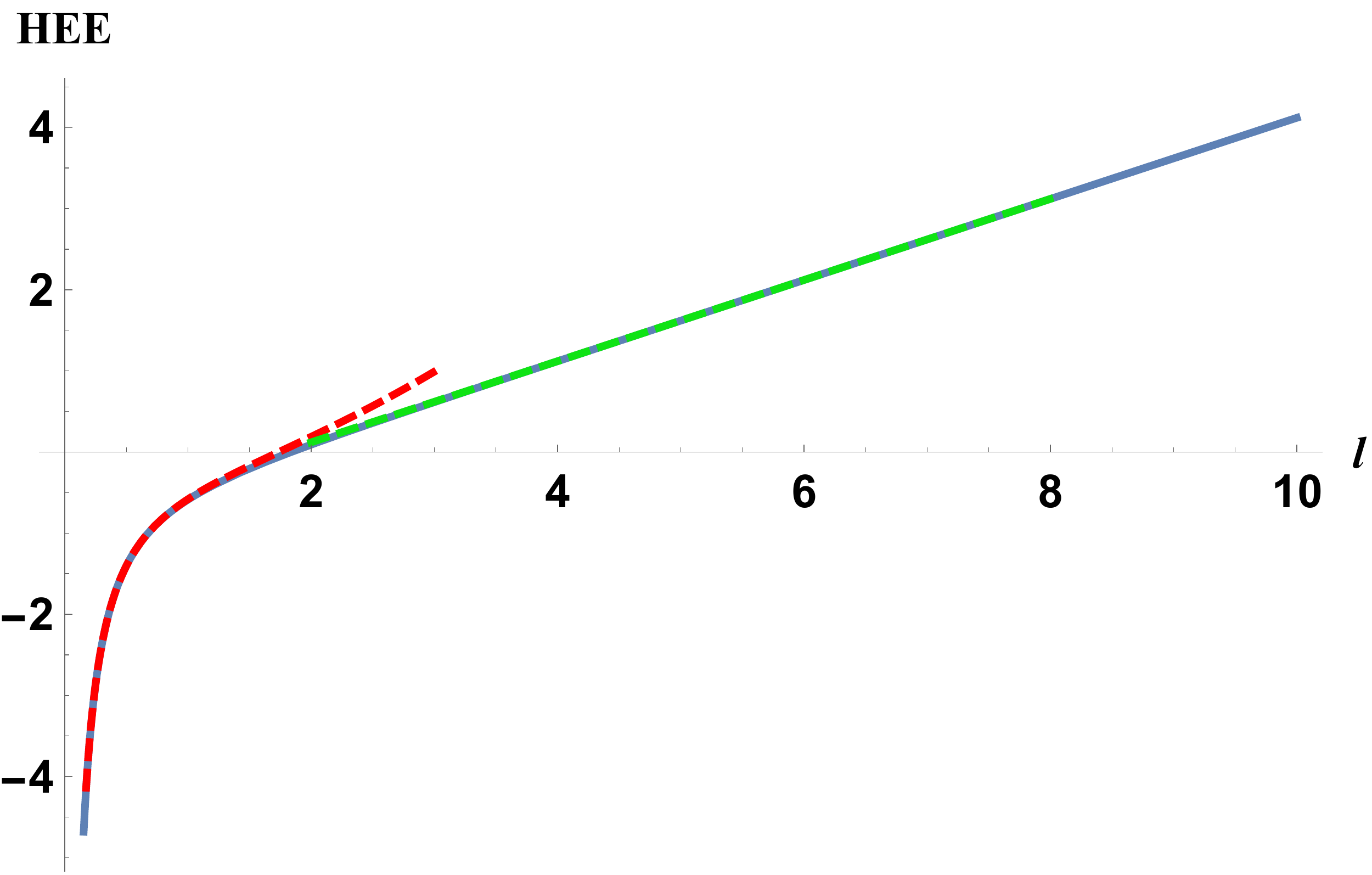}
		\caption{\emph{Left plot}: Comparison of numeric HEE (blue) with semi-analytic HEE for $\ell \ll {\rho _H}$ (red) in AdS$_3$-BH. \emph{Right plot}: Comparison of numeric HEE (blue) with semi-analytic HEE for $\ell \ll {\rho _H}$ (red) and $\ell \gg {\rho _H}$ (green) limits in AdS$_4$-BH.}\label{comparison}
	\end{figure}

	In this section, we study the holographic $n$-partite information in AdS-BH geometry focusing on four and five partite information. Making use of the metric (\ref{bulk}) and equations (\ref{npar}) and (\ref{a3-2}), one can obtain the relevant HEE in this background as follows
	\begin{equation}\label{heebh}
		S = \frac{{{{\rm{L}}^{d - 2}}}}{{4{G_N}}}\int_\epsilon ^{{\rho _t}}
		{d\rho \frac{1}{{{\rho ^{d - 1}}\sqrt {f(\rho )\left( {1 - {{\left( {\frac{\rho }{{{\rho _t}}}} \right)}^{2(d - 1)}}} \right)} }}},
	\end{equation}
	where $\rho_t$ is the turning point of the hypersurface in the bulk. In general there is no analytic solution and the numerical calculation is actually needed to carry out the integrals in (\ref{a3-1}) and (\ref{heebh}). However, for some special limits one can obtain semi-analytic expressions for HEE as follows (see Appendix)
	\begin{equation}
		{S_{\text{BH}}}(\ell)=
		\begin{cases}
			{S_{\text{vac}}}\left( \ell \right) + \frac{{{L^{d - 2}}}}{{4{G_N}}}{c_1}\left( d \right)\rho_H^{-d}{\ell^2}, & \quad \ell \ll {\rho _H}  \\
			\frac{{{L^{d - 2}}}}{{4{G_N}}}\left( {\frac{1}{{\left( {d - 2} \right){\varepsilon ^{d - 2}}}} +
				\frac{\ell}{{2{\rho _H}^{d - 1}}} - \frac{{{c_2}(d)}}{{{\rho _H}^{d - 2}}}} \right), & \quad \ell \gg {\rho _H}$\,\,\,$ {\rm{and}} $\,\,\,$ d > 2  \\
		\end{cases}\label{sbh}
	\end{equation}
	where ${S_{\text{vac}}}$ stands for the entanglement entropy for the AdS which is defined by \eqref{svac} and $c_1$ is given by
\[{c_1}\left( d \right) = \frac{1}{{16(d + 1)\sqrt \pi  }}\frac{{\Gamma {{\left( {\frac{1}{{2d - 2}}} \right)}^2}\Gamma \left( {\frac{1}{{d - 1}}} \right)}}{{\Gamma {{\left( {\frac{d}{{2d - 2}}} \right)}^2}\Gamma \left( {\frac{1}{2} + \frac{1}{{d - 1}}} \right)}}.\]
Note that ${c_2}(d)$ is a constant that depends on $d$, \emph{e.g.} ${c_2}(3) = 0.88$ and $c_2(4)=0.33$. Here we use the numerical methods to
	compute the integrals for $0<\rho _t<\rho _H$ with step of $0.00005$ and $\epsilon = 0.0001$  which results in lists of values for $\ell\left(\rho _t\right)$ and ${S_{BH}}\left(\rho _t\right)$. The relevant HEE is obtained in terms of the entangling regions length by interpolating the list of $\{\ell\left(\rho _t\right),{S_{BH}}\left(\rho _t\right)\}$. Fig.\ref{comparison} shows that numerical computations are in good agreement with the semi-analytic ones.
	
\begin{figure}[h!]\label{fig1}
		\centering
		\includegraphics[width=6.5cm]{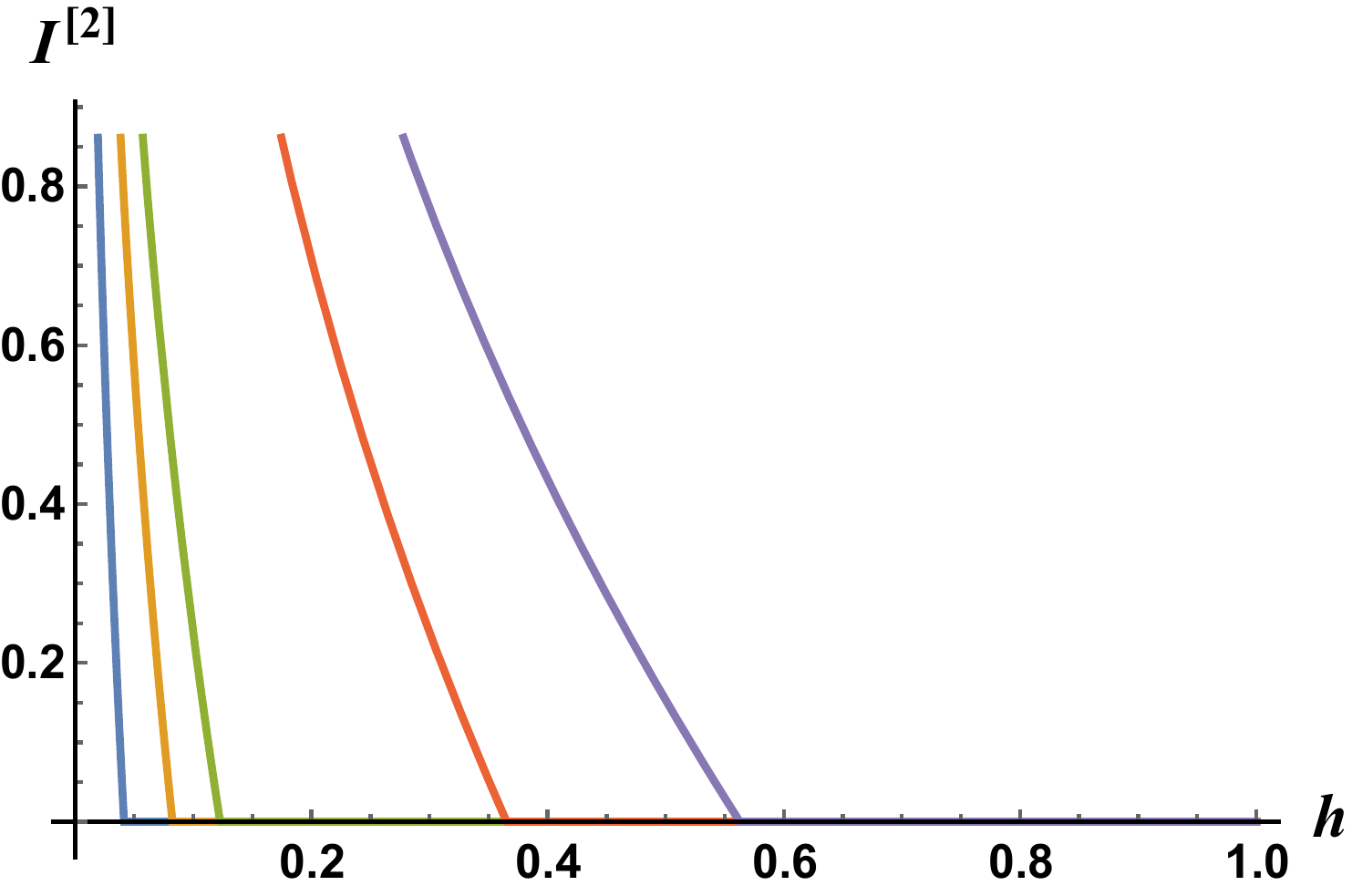}\,\,\,\,\,\,\,\includegraphics[width=6.5cm]{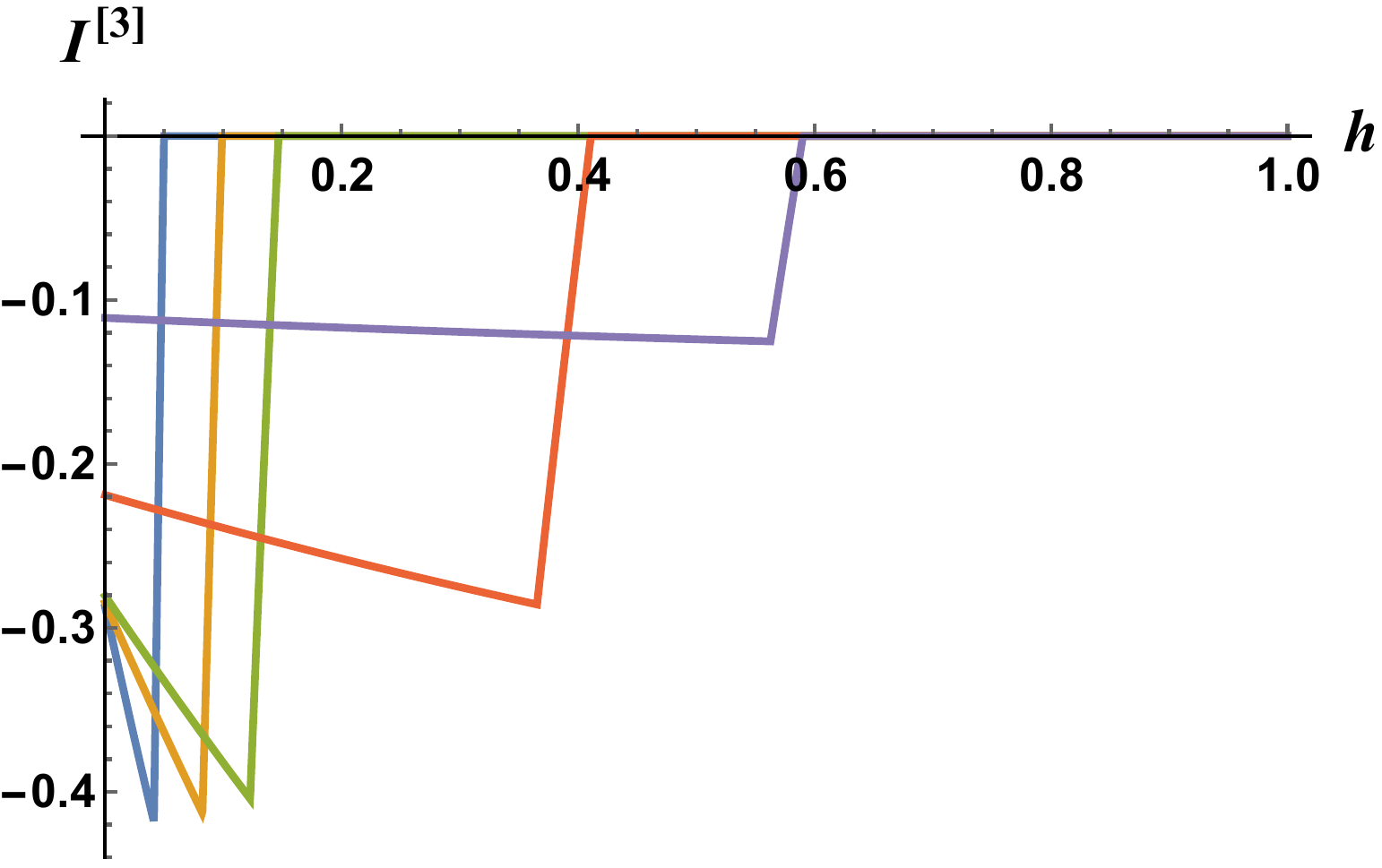}
		\includegraphics[width=6cm]{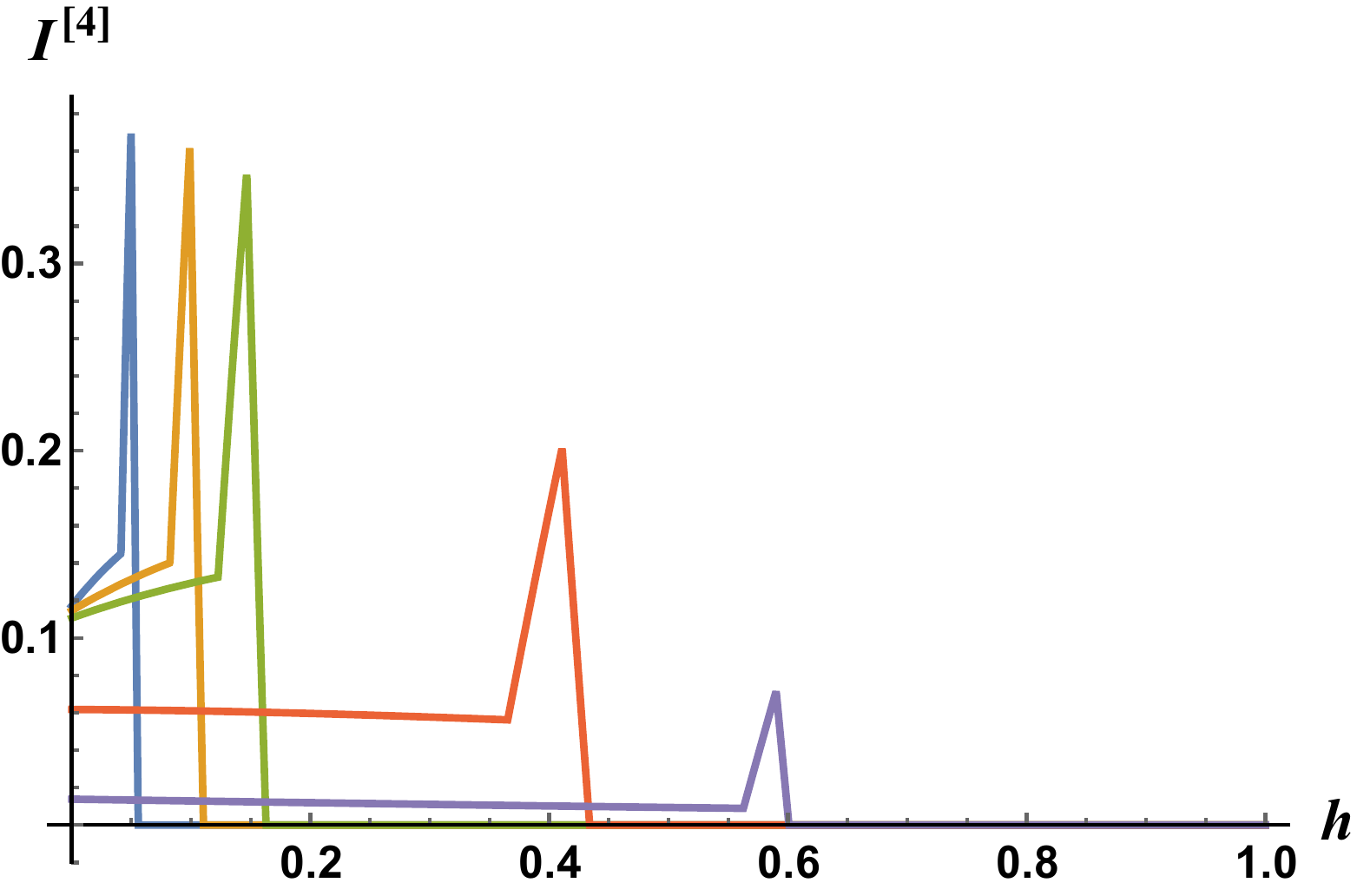}\,\,\,\,\,\,\,\includegraphics[width=6cm]{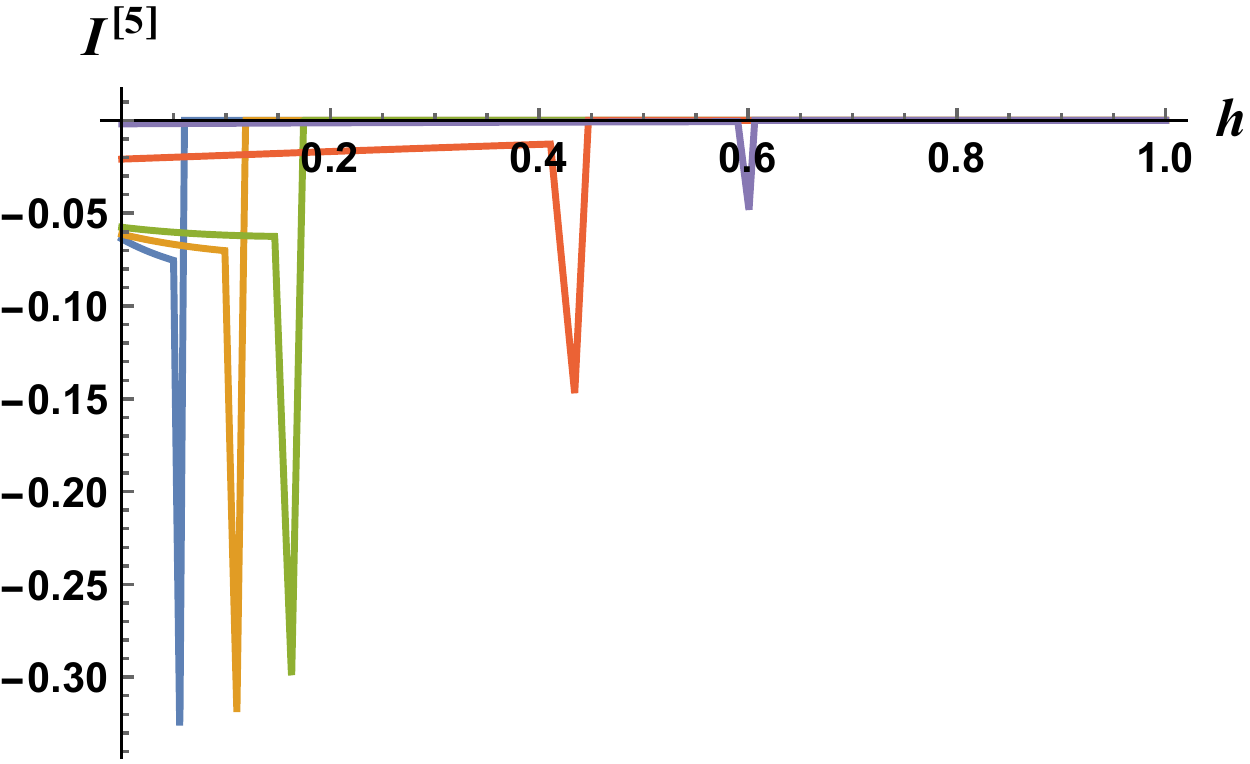}
		\caption{ Holographic $n$-partite information in AdS$_3$-BH for $\ell = 0.1,{\rm{ }}0.2,{\rm{ }}0.3,{\rm{ }}1,{\rm{ }}2$ from left to right. }\label{bh}
	\end{figure}

\begin{figure}[h!]
		\centering
		\includegraphics[width=6.5cm]{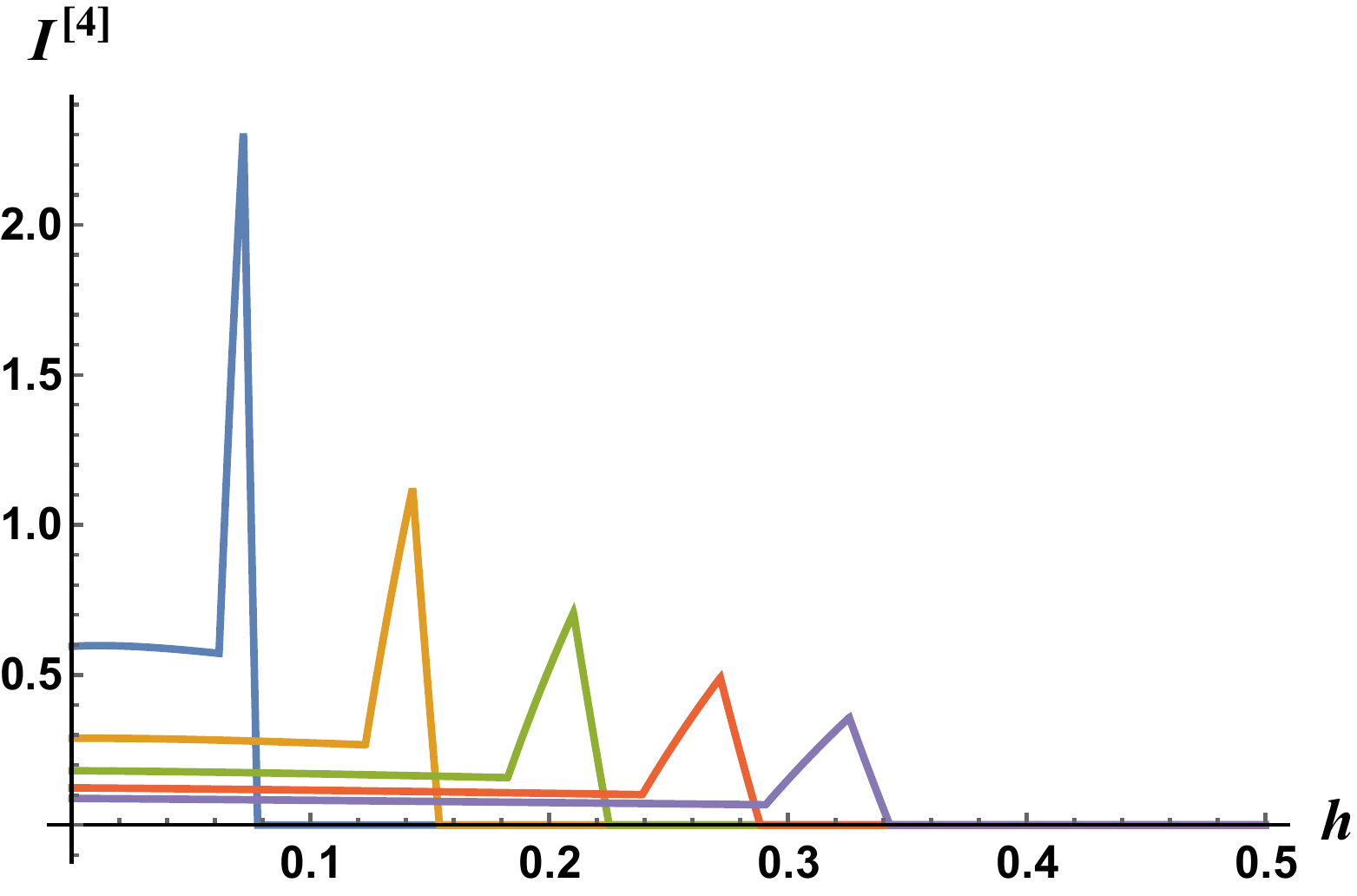}
		\caption{Holographic $n$-partite information in AdS$_4$-BH for $\ell = 0.1,{\rm{ }}0.2,{\rm{ }}0.3,{\rm{ }}0.4,{\rm{ }}0.5$ from left to right.}\label{bh1}
	\end{figure}

	Having obtained the HEE in this background, it is straightforward to study the holographic $n$-partite information via (\ref{npar}). Holographically the main step is to compute the minimal surface among all possible configurations. The strategy of finding such surface is similar to what presented in the previous section. Therefore, we only present the results of numeric computation, mentioning that according to different ranges of variables in which the sign of holographic $n$-partite information is investigated, the minimal configurations must be chosen. By plotting the resulting holographic four and five partite information for various values of parameters, it is verified that the sign is fixed in AdS-BH geometry; this is depicted in Fig.s (\ref{bh}, \ref{bh1}). \\ In AdS$_4$ and AdS$_4$-BH, holographic 5-partite information seems to take either sign and we leave the question of other possible relations as a challenge for the future.
	
	\section{Discussions}
	
	In Ref.\cite{Hayden:2011ag}, assuming the RT formula in quantum field theories
	with holographic duals, it was shown that the mutual informations among arbitrary disjoint spatial regions
	$A$, $B$ and $C$ obey the inequality $I(A: B\cup C)\geq I(A:B)+I(A:C)$. This inequality indicates that the mutual information in monogamous. The monogamy relation is actually characteristic of measures of quantum entanglement which
	suggests that correlations in holographic theories arise primarily from entanglement rather than classical correlations. In a general quantum field theory, the mutual information does not obey the monogamy condition, because it encodes both entanglement and classical correlations. Therefore, the property of monogamous is only valid in holographic theories which suggests that in such theories quantum entanglement becomes significant and dominates over classical correlations.\\
	In this paper we have mainly studied the monogamy relations for $n$-partite information by employing the holographic prescription for computing entanglement entropy. Specifically, we have considered the $n=4$ and $n=5$ cases in a vacuum and an excited state corresponding to pure AdS and AdS-BH geometries. We have chosen parallel infinite strips as the entangling regions in a constant time slice. In particular, in order to simplify the computations, we considered the equal width strips ($\ell_i=\ell$) with the same spatial separation ($h_i=h$). The main difficulty that we encountered was to find the minimal area configuration among the different RT surfaces corresponding to boundary entangling regions. Actually our assumption on the equality of the entangling regions width and separation between them reduces the number of possible configurations. The competition between these configurations depends on the ratio $r\equiv\frac{h}{\ell}$. In this set-up, our main results can be summarized as follows.\\
	In the case of four disjoint parallel strips with same width $\ell$ separated by same distance $h$, we have shown that 4-partite information is always positive for both vacuum and thermal state. Although we have performed a numerical study for various lengths and dimensions, it seems that these results hold in general. Also, we have investigated the possible phase transitions of 4-partite information and we have found the critical points $(r_i)$ where the corresponding minimal configurations, \emph{i.e.} RT  surfaces, have been changed. On the other hand for five disjoint parallel strips with same width $\ell$ separated by same distance $h$, we have found that 5-partite information is always negative in the vacuum and thermal state in three dimensional background. Again in this case, we discussed about the possible critical points and phase transitions. Making use of equation \eqref{npar2} which gives the 5-partite information in terms of 4-partite information and the latter result, \emph{i.e.} $I^{[5]}<0$ for CFT$_2$, it is indicated that in the present set-up of parallel strips with same width and separation, the holographic 4-partite information is monogamous in three dimensional AdS background. 
	
	There are several directions which one can follow to further investigate this study. One of the main directions is to consider the time dependent backgrounds and check how the sign of these quantities changes during the thermalization process. Another generalization is to investigate the role of Lifshitz and hyperscaling violating exponents on these monogamy relations. We leave further investigations to future works.
	
	\subsection*{Acknowledgments}
	
	We would like to thank M. Alishahiha and M. Reza Mohammadi Mozaffar for their useful comments and discussions. MRT also wishes to thank A. Mollabashi, F. Omidi and A. Faraji for some related discussions. This work has been supported in parts by Islamic Azad University Central Tehran Branch.
	
	\section*{Appendix}
	\appendix
	\section{Holographic entanglement entropy: A short review} \label{App}
	
	In this appendix, we will recall the holographic computation of the entanglement entropy whose dual gravity is an AdS geometry given by (\ref{bulk}). In order to compute the entanglement entropy of a system described by (\ref{strip}) in the holographic context, it is needed to find a co-dimension two surface with minimum area in the bulk regarding the point that at the boundary this surface coincides with that of the entangling region. To find the minimal surface in the bulk the profile can be parameterized by $x_{1}=x(\rho)$, so that the induced metric on this profile is obtained as follows
	\begin{equation}\label{indmetric}
		d{s^2}_{\text{ind}} = \frac{1}{{{\rho ^2}}}\left( {\left( {f{{(\rho )}^{ - 1}} + x'{{(\rho )}^2}} \right)d{\rho ^2} + \sum\limits_{i = 2}^{d - 1} {d{x_i}^2} } \right)
	\end{equation}
	in which $x'(\rho ) = \frac{{\partial x(\rho ){\rm{ }}}}{{\partial
			\rho }}$.
	Consequently, the area functional of the co-dimension two hypersurface in the bulk becomes
	\begin{equation}\label{areafunctional}
		A = \frac{{{{\rm{L}}^{d - 2}}}}{2}\int {\frac{{\sqrt {f{{(\rho )}^{ - 1}} + x'{{(\rho )}^2}} }}{{{\rho ^{d - 1}}}}d\rho }  \equiv \frac{{{{\rm{L}}^{d - 2}}}}{2}\int {{\mathcal{L}} d\rho } .
	\end{equation}
	Making use of the variational principle, one can minimize the proposed surface. Since the integrand is independent of $x$, the corresponding momentum is a constant of motion, \emph{i.e.}
	\begin{equation}\label{momentum1}
		P_{x}=\frac{\partial \mathcal{L}}{\partial x'}={\rm{constant}}.
	\end{equation}
	By substituting $\mathcal{L}$ from (\ref{areafunctional}) in (\ref{momentum1}) and doing some simplification, one arrives to
	\begin{equation}\label{momentum2}
		x'(\rho ) = \frac{1}{{\sqrt {f(\rho )\left( {\frac{1}{{{P_x}^2{\rho ^{2(d - 1)}}}} - 1} \right)} }}.
	\end{equation}
	Note that at the turning point of hypersurface in the bulk, ${\rho _t}$, one has $x'({\rho _t}) = \infty$. By applying this condition to (\ref{momentum2}) and after doing some calculations one gains
	\begin{equation}\label{momentum3}
		{P_x}^2 = \frac{1}{{{\rho _t}^{2(d - 1)}}}.
	\end{equation}
	Replacing (\ref{momentum3}) in (\ref{momentum2}) results in
	\begin{equation}\label{xprime}
		x'(\rho)=\frac{\left(\frac{\rho}{\rho_{t}}\right)^{(d-1)}}{\sqrt{f(\rho)\left(1-\left(\frac{\rho}{\rho_{t}}\right)^{2(d-1)}\right)}}.
	\end{equation}
	By integrating from the above formula one obtains
	\begin{equation}\label{a3-1}
		\frac{\ell}{2} = \int_0^{{\rho _t}} {d\rho \frac{{{{\left(
							{\frac{\rho }{{{\rho _t}}}} \right)}^{(d - 1)}}}}{{\sqrt {f(\rho )
						\left( {1 - {{\left( {\frac{\rho }{{{\rho _t}}}} \right)}^{2(d -
									1)}}} \right)} }}}.
	\end{equation}
	With the help of (\ref{xprime}), it is straightforward to obtain
	\begin{equation}\label{a3-2}
		A = {{\rm{L}}^{d - 2}}\int_\epsilon ^{{\rho _t}} {d\rho \frac{1}{{{\rho ^{d - 1}}\sqrt {f(\rho )
						\left( {1 - {{\left( {\frac{\rho }{{{\rho _t}}}} \right)}^{2(d - 1)}}} \right)} }}}
	\end{equation}
	where $\epsilon$ stands for the UV cut-off. After making use of (\ref{ee}), one can obtain the expression of HEE.
	
	It is noted that in general there is no analytic solution for the above equation, however, one can find solutions in the case of AdS geometry with $f(\rho ) = 1$, which is dual to the vacuum state of the underlying CFT. In this case after making use of (\ref{a3-1}), the turning point $\rho _t$ can be written in terms of $\ell$ as follows
\begin{equation}\label{turningpoint}
{\rho _t} = \frac{{\Gamma \left( {\frac{1}{{2(d - 1)}}} \right)}}{{2\sqrt \pi  \Gamma \left( {\frac{d}{{2(d - 1)}}} \right)}}\ell.
\end{equation}
Therefore, the holographic entanglement entropy in AdS background
	is obtained as \eqref{svac}.
	
For $f(\rho)\neq0$, in general, there is no explicit analytic expression for the
entanglement entropy, however, in some limits one can find a semi-analytic solution as stated below. 
\begin{itemize}
	\item $\ell\ll \rho_H$ limit: \\In this limit one can expand the area \eqref{a3-2} around the
$f=1$ which leads to the following expression 
\begin{equation}
\Delta {A}=\frac{L^{d-2}}{2}\int d\rho\;\; \delta_f\left(\frac{\sqrt{{f}^{-1}+x'^2}}{\rho^{d-1}}\right)\bigg|_{f=1}\Delta f.
\end{equation}
Therefore one can write the entanglement entropy as follows
\begin{equation}\label{SBH0}
S_{\text{BH}}=S_{\text{vac}}+\frac{L^{d-2}}{4G_N} c_1\rho_H^{-d}\ell^2,
\end{equation}
where
\begin{equation}
c_1=\frac{1}{16(d+1)\sqrt{\pi}}\;\frac{\Gamma(\frac{1}{2(d-1)})^2\Gamma(\frac{1}{d-1})}{\Gamma(\frac{d}{2(d-1)})^2\Gamma(\frac{1}{2}+\frac{1}{d-1})},
\end{equation}
and $S_{\rm vac}$ stands for the entanglement entropy of the vacuum solution given by \eqref{svac}.
\item  $\ell\gg\rho_H$ limit:\\
In this limit one can show that $\rho_t\backsimeq \rho_H$ meaning that the minimal surface
is extended all the way to the horizon. Therefore, one can expand \eqref{a3-1} and \eqref{a3-2} in this limit which results in 
\begin{equation}
\frac{\ell}{2}\approx\rho_H \int_0^{1} \frac{\xi^{d-1}d\xi}{
	\sqrt{(1-\xi^d)\left(1-\xi^{2(d-1)}\right)}},\;\;\;\;\;\;\;\;\;
A\approx\frac{L^{d-2}}{\rho_H^{d-2}}\int_{\frac{\epsilon}{\rho_H}}^{1} \frac{d\xi}{\xi^{d-1}
	\sqrt{(1-\xi^d) \left(1-\xi^{2(d-1)}\right)}}.
\end{equation}
In the expression of area, because of the double zero in the square roots, $\xi=1$ point gives the main contributions. After expanding the area around $\xi=1$ and doing some algebra, one receives 
\begin{equation}
A\approx\frac{L^{d-2}}{\rho_H^{d-1} }\;\frac{\ell}{2}
+\frac{L^{d-2}}{\rho_H^{d-2} }
\int_{\frac{\epsilon}{\rho_H}}^{1} d\xi\; \frac{\sqrt{1-\xi^{2(d-1)}}}{\xi^{d-1}
	\sqrt{1-\xi^d }}.
\end{equation}
The second integral can be solved numerically leading to $\frac{1}{(d-2)\epsilon^{d-2}}-c_2(d)$,
where $c_2(d)$ is a positive number \cite{Alishahiha:2014jxa}. Finally the holographic entanglement entropy in this limit becomes 
\begin{equation}\label{SBH1}
S_{\rm BH}=\frac{A}{4G_N}\approx\frac{L^{d-2}}{4G_N }\left(\frac{1}{(d-2)\epsilon^{d-2}}+\frac{\ell}{2\rho_H^{d-1}}
-\;\frac{c_2(d)}{\rho_H^{d-2}}\right).
\end{equation}
\end{itemize}
	

\end{document}